\documentclass[review]{elsarticle}

\usepackage{amsthm,amsmath} 
\usepackage{amssymb}
\usepackage{xcolor}
\usepackage{algorithm} 
\usepackage{algorithmic}
\usepackage{pstool}
\usepackage{subcaption}
\usepackage{enumitem}
\usepackage{amsmath, amsfonts, mathtools}
\usepackage{algorithm,algorithmic}
\usepackage{natbib}
\usepackage{hyperref}
\hypersetup{
    colorlinks = true,
    citecolor = {blue},
}
\usepackage{hypernat}

\graphicspath{{figures/}}

\def\1{{\mathbf 1}}

\newcommand{\s}{\ensuremath{\mathbb{S}}}

\numberwithin{equation}{section}
\hypersetup{pdfauthor = {J. Derek Tucker, Lyndsay Shand, and Kenny Chowdhary}, pdftitle = {Multimodal Bayesian Registration of Noisy Functions using Hamiltonian Monte Carlo}, pdfsubject = {Computational Statistics}, pdfkeywords = {amplitude variability, function alignment, phase variability}, pdfcreator = {LaTeX with hyperref package}, pdfproducer = {xelatex}}

\journal{Computational Statistics \& Data Analysis}
\begin{document} 

\tabcolsep 2pt 
\begin{frontmatter}
\title{Multimodal Bayesian Registration of Noisy Functions using Hamiltonian Monte Carlo}

\author[snl]{J. Derek Tucker\corref{cor1}}
\ead{jdtuck@sandia.gov}
\author[snl]{Lyndsay Shand}
\author[snl]{Kenny Chowdhary}
\cortext[cor1]{Corresponding author. Tel.: +1 505 284 8415}	
\address[snl]{Sandia National Laboratories, PO Box 5800 MS 0829, Albuquerque, NM 87185} 

\begin{abstract}
  Functional data registration is a necessary processing step for many applications. 
  The observed data can be inherently noisy, often due to measurement error or natural process uncertainty; which most functional alignment methods cannot handle. 
  A pair of functions can also have multiple optimal alignment solutions, which is not addressed in current literature. 
  In this paper, a flexible Bayesian approach to functional alignment is presented, which appropriately accounts for noise in the data without any pre-smoothing required.  
  Additionally, by running parallel MCMC chains, the method can account for multiple optimal alignments via the multi-modal posterior distribution of the warping functions. 
  To most efficiently sample the warping functions, the approach relies on a modification of the standard Hamiltonian Monte Carlo to be well-defined on the infinite-dimensional Hilbert space. 
  This flexible Bayesian alignment method is applied to both simulated data and real data sets to show its efficiency in handling noisy functions and successfully accounting for multiple optimal alignments in the posterior; characterizing the uncertainty surrounding the warping functions.
\end{abstract}

\end{frontmatter}

\section{Introduction}\label{sec:intro}

Functional registration or time-warping, refers to the process of aligning two or more functions or curves in time and is often a pre-processing step necessary for appropriately analyzing such functions.
By registering functions before performing statistical analysis on such functions, we can account for arbitrary reparameterizations of functions. 
This can lead to incorrect analyses of many functional data applications, such as the well-known growth rate curve analysis.
Function registration is heavily relied upon in image and shape registration where analysis results can vastly differ depending on whether the data has been properly registered or not.
More details on the importance of functional data registration can be found in \cite{FSDA, marron2015}, and \cite{ramsay2005}. 
Functional data registration is also referred to as phase-amplitude separation, because the underlying goal of the procedure is to effectively distinguish phase ($x$-axis) from amplitude ($y$-axis) variation. For example, in tracking the migration paths of birds or hurricanes, functional data registration would allow one to isolate the statistical variation in the paths, from the statistical variation in the speed of traversing the paths.
The registration of a pair of functions results in an optimal warping function, which enables one function to be aligned to the other. The resulting aligned functions characterize the amplitude variability, while the warping function captures the phase variability between the two functions.

Early approaches to functional data alignment fall short in four major ways. 
(1) They lack the ability to characterize the uncertainty of the optimal warping function.
(2) They assume the observed functions are naturally smooth and lack measurement error. 
(3) They do not consider more than one optimal alignment between a pair of functions.
Finally, (4) Most methods do not use a proper distance as the objective function for registration. 

Traditional approaches to functional data alignment, which are demonstrated in \cite{srivastava-etal-JASA:2011,ramsay2005, sangalli-etal2:2010}, and \cite{kneip-ramsay:2008}, are not flexible enough to characterize the uncertainty of the resulting optimal alignment solution.
These approaches make the common assumption that the functions to be aligned are smooth, and thus they break down when a function exhibits too much noise. 
These methods usually rely on a derivative which is extremely noisy when the original functions are noisy, presenting additional computational challenges.

Recently, Bayesian frameworks have been proposed that can characterize the uncertainty of the warping function solution \citep{telesca2008,kurtek2017,lu2017, cheng2016}.
The clear difference between these approaches is in how they specify a prior on the space of the warping function; as it lives on a nonlinear infinite dimensional manifold.
The more recent Bayesian approaches by \cite{cheng2016, lu2017}, and \cite{kurtek2017} rely on the square root velocity function (SRVF) representation of the warping function space, introduced by \cite{srivastava2011} and \cite{srivastava-etal-JASA:2011}, to simplify the complicated geometry. We will also take advantage of this transformation and provide more details in Section \ref{sec:review}. Although these approaches demonstrate how a Bayesian framework can naturally handle uncertainty quantification of the warping function through posterior inference, few have account for measurement error in the observed data and none allow for a multimodal posterior, i.e. multiple optimal warpings.

Only the most recently proposed approach \cite{matuk2019}, has addressed the need to account for observed measurement error. 
There approach is for sparsely sampled functions where they propose a data-driven prior for inference and solely look at sparsely sampled functions and do not treat the dense problem.
Previous approaches, Bayesian or otherwise, have assumed the observed functions to be aligned are naturally smooth. Although smoothness is a convenient assumption, it can be an invalid one in many practical applications.
Recent work \cite{panaretos:2016}, utilized a fPCA construction to avoid overfitting when the data has additive amplitude variation. 
Wrobel \cite{wrobel:2019}, recently extended the alignment to non-Gaussian data. Both of these methods do not address the estimation of the additive noise process that we will address in this paper.
Lastly, no current approaches consider the challenging case of multiple optimal alignments between a pair of functions; or briefly mention a possible solution without application. 

In this paper, we propose a hierarchical Bayesian approach to the registration of a pair of noisy functions on $\mathbb{R}^1$ and demonstrate its advantages over previously proposed registration algorithms.
We propose a new framework for functional alignment which relies on a proper distance metric, is robust to noise, capable of characterizing the uncertainty of the warping function, and can account for multiple optimal alignments.
The novelty of our new framework is the following: 
Most notably, we employ parallel MCMC chains to capture multi-modal posterior distributions of our warping functions; to reflect the case when there are multiple optimal alignments.
Additionally, we favor the geometric Hamiltonian Monte Carlo algorithm of \cite{beskos2017} ($\infty$-HMC), which is nontrivial to apply to problem of functional data alignment.  
The implementation of $\infty$-HMC within our MCMC sampler can appropriately account for the complex geometry of the target space and more efficiently sample the warping function.
Lastly, our approach accounts for  appropriate propagates measurement error, a challenge which has only recently been addressed by \cite{matuk2019}.

This paper is organized in the following way. Section \ref{sec:review} reviews pairwise functional registration in $\mathbb{R}^1$, the general challenges associated with it, and introduces the square-root velocity function representation and proper distance metric relied upon throughout this paper.
Section \ref{sec:pairwise} specifies our hierarchical Bayesian model for registering a pair of noisy functions on $\mathbb{R}^1$, including our MCMC algorithm and multi-chain approach to the challenge of multiple alignments.
In Section \ref{sec:examples}, we evaluate the approach on simulated functional data as well as two real data sets: a SONAR dataset that is naturally noisy and iPhone-collected accelerometer data with multiple possible alignments.
Code for the method is found in the \verb+fdasrvf+ Matlab package on GitHub\footnote{\url{https://github.com/jdtuck/fdasrvf_MATLAB}}.
Finally, we discuss the impact of this approach, extensions to multiple pairwise alignments, nontrivial extensions to functions on more complex geometries, and future work in Section \ref{sec:disc}.

\section{A review of function registration in $\mathbb{R}^1$}\label{sec:review}

Following the notation of \cite{tucker2013} and without loss of generality, let $f$ be a real-valued and absolutely continuous function on domain $[0, 1]$.
Note that in practice, $f$ is observed as discrete and interpolation can be used to more easily perform the requisite calculations.
Let $\mathbb{F}$ denote the set of all such functions and $\Gamma$ denote the set of boundary-preserving diffeomorphisms: $\Gamma = \{\gamma : [0,1]\mapsto [0,1] \mid \gamma(0)=0,\gamma(1)=1\}$ such that the mapping $ [0,1]\mapsto [0,1]$ is bijective (invertible) and differentiable.
From this point on, we will refer to $\Gamma$ as the set of warping functions.
Then, for any $f\in \mathbb{F}$ and any $\gamma \in \Gamma,$ $f \circ \gamma$ denotes the time-warping of $f$ by $\Gamma$.

For simplicity, consider the pairwise alignment problem where we wish to align functions $f_1,f_2 \in \mathbb{F}.$ A simplified registration problem can be formulated by finding a warping, $\gamma^*,$ that minimizes the cost of translating $f_2$ to $f_1$, for chosen cost function
$$\gamma^* = \text{argmin}_{\gamma} \| f_1 - f_2 \circ \gamma\|^*, $$
where $\| \|^*$ is our distance metric yet to be chosen.\footnote{Although we are proposing a more robust approach, we are introducing a simplified registration problem to motivate the use of a proper distance metric.}
There are a few main challenges that all alignment approaches face and handle differently.
One challenge of function alignment, as pointed out by \cite{srivastava-etal-JASA:2011} and \cite{tucker2013}, is finding a cost function to do the alignment that is both symmetric, that is, aligning $f_1$ to $f_2$ is the same as aligning $f_2$ to $f_1$,  and positive-definite, so that the metric is always non-negative and zero if and only if $f_1$ and $f_2$ are the same function after alignment.
A natural choice is the usual $\mathbb{L}^2$ norm, denoted by $\|\cdot\|^2$, but it does not satisfy the symmetry requirement. 
More precisely, 
$$\text{argmin}_{\gamma}\|f_1 \circ \gamma - f_2 \|^2 \neq \text{argmin}_{\gamma}\|f_1 - f_2 \circ \gamma \|^2. $$
Second, there can be the issue of degeneracy in which case $\gamma^*$ is so distorted it can align functions which should not be aligned; giving a false impression that these functions are \textit{close} in the $\mathbb{L}^2$ sense when indeed they are not.
Lastly, the $\mathbb{L}^2$ norm is not invariant under warping. That is, 
$$\| f_1 - f_2 \|^2 \neq \| f_1 \circ \gamma - f_2 \circ \gamma\|^2.$$
This means that two functions which are similar in an $\mathbb{L}^2$ sense, could be vastly different under the same warping, and vice versa.

To overcome these three challenges, we follow \cite{tucker2013}, \cite{cheng2016} and \cite{lu2017} and take advantage of the square-root velocity function (SRVF) representation introduced by \cite{srivastava2011}.
A function $f\in \mathbb{F}$ can be represented as a SRVF via the mapping:
$$q:[0,1]\mapsto \mathbb{R},\ \  q(f(t))=\hbox{sign}(\dot{f}(t))\sqrt{|\dot{f}(t)|} \hbox{ for any } f\in \mathbb{F}.$$
There is an equivalency, up to a constant, between the SRVF of $f$ and $f$ itself and it is given by $f(t) = f(0) + \int_0^t q(f(t))|q(f(t))| ds$, where $|\cdot|$ is the absolute value. Moreover, the SRVF of $f \circ \gamma$ is $(q \circ \gamma)\sqrt{\dot{\gamma}}$.
The SRVF transformation lends itself naturally to the Fisher-Rao (FR) metric. It can be shown that the FR distance between two functions is equivalent to the $\mathbb{L}^2$ distance between their respective SRVF transformations, i.e., $\| f_1 - f_2 \|_{FR} = \|q(f_1) - q(f_2) \|^2$ (see \cite{srivastava-etal-JASA:2011}). More importantly, the FR metric is phase invariant under warping, i.e., $\| f_1 \circ \gamma - f_2 \circ \gamma \|_{FR} = \|f_1 - f_2 \|_{FR}$ and \cite{srivastava2011} shows that the FR metric is symmetric under warping as well, i.e., $\| f_1 \circ \gamma - f_2 \|_{FR} = \| f_1 - f_2 \circ \gamma \|_{FR}$.

Following this framework and without loss of generality, we will align the SRVF of $f_2$ to the SRVF of $f_1$, and then map the aligned functions back to the original space $\mathbb{F}$.
In other words, we seek to estimate the warping function that minimizes $||q_1- (q_2,\gamma)||^2$, where $(q_2,\gamma)=(q_2\circ\gamma)\sqrt{\dot{\gamma}}$ and $q_j \doteq q(f_j(t))$ for $j=1,2$.

Additional challenges in function alignment come when placing the problem in a Bayesian framework, namely selecting a prior distribution for $\gamma \in \Gamma$. Recall that $\Gamma$ is the space of diffeomorphic function mappings from $[0,1] \mapsto [0,1]$. 
Both optimization and Bayesian inference over this space is difficult, in particular, because this space is nonlinear and infinite dimensional. 
For example, the sum or scalar product of functions in $\Gamma$ is not necessarily still contained in $\Gamma$. 
To overcome this difficulty, we follow the approach of \cite{lu2017,tucker2013} which transforms $\gamma$ to its corresponding SRVF representation, and exploits the Riemannian-geometric structure of this transformation.
Which ultimately allows us to utilize more traditional inference or optimization algorithms that rely on the linearity of the underlying search space.

\begin{figure}
\centering
\includegraphics[scale=0.3]{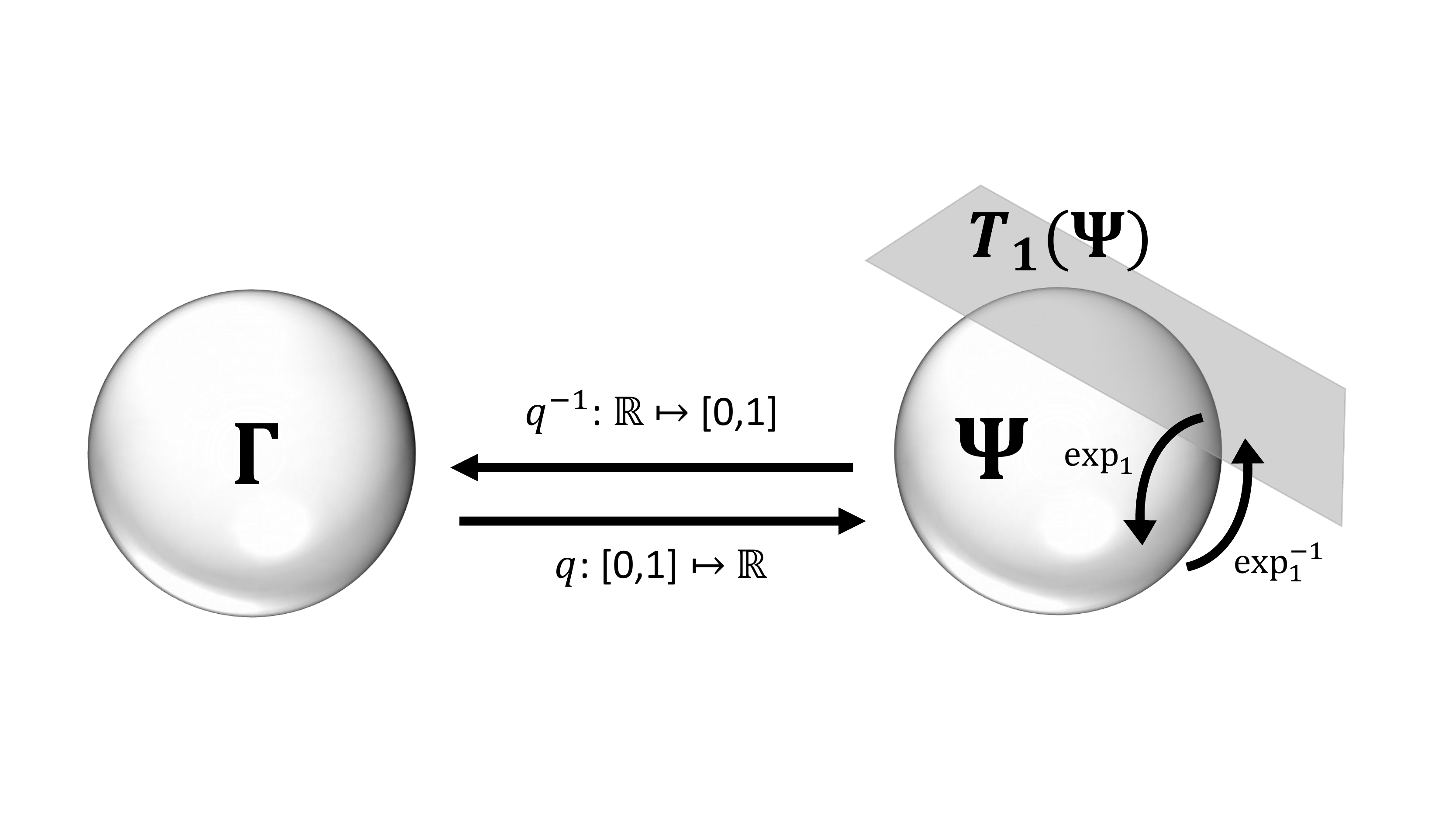}
\caption{A summary of the mapping from $\Gamma$ to $T_1(\Psi)$ where the space $\Gamma$ is a nonlinear manifold and $\Psi$ is the unit sphere in the Hilbert space $\mathbb{L}^2([0,1])$.}\label{fig:mappings}
\end{figure}

To understand how this transformation works, it is helpful to think of the set of warping functions $\Gamma$ as the space of univariate cumulative distribution functions for random variables on $[0,1]$. Then, for each $\gamma \in \Gamma$ there is an associated density function. 
Now, let $\psi=\sqrt{\dot{\gamma}}$ be the corresponding SRVF of $\gamma$ and $\Psi^+=\{\psi: [0,1]\mapsto \mathbb{R}^+\mid||\psi||^2=1\}$ \citep{bhattachayya1943} be the space of square-root densities (SRD). 
The SRD space is the positive orthant of the unit sphere in the Hilbert space $\mathbb{L}^2([0,1])$ denoted by $\Psi\{\psi: [0,1]\mapsto \mathbb{R}\mid||\psi||^2=1\}$. 
While this space is still infinite and nonlinear, it is much more simply defined. Since the SRVF is a bijective mapping and $\gamma(0)=0,$ we can reconstruct $\gamma$ by the inverse mapping $q^{-1}(\psi)(t)=\int_0^t\psi^2(s)ds.$
We can further simplify, and even linearize $\Psi^+$ by mapping from the top half of the unit sphere onto a tangent space at $\psi$ defined as
$$T_{\psi}(\Psi)=\{g\in \mathbb{L}^2\mid\int_0^1g(s)\psi(s) ds=0\}.$$
For simplicity, we typically take $\psi = 1$, i.e., the identity function. 
Geometric details on $T_1(\Psi)$ can be found in \cite{srivastava2007} and \cite{kurtek2012,tucker2013}.
The exponential map and its inverse can be used to map between $\Psi$ and $T_1(\Psi):$

$$\begin{aligned}
&\exp_1: & T_1(\Psi)\mapsto \Psi &\hspace{4mm} \exp_1(g)=\cos(||g||)+\sin(||g||)\frac{g}{||g||},~~  g\in T_1(\Psi)\\
&\exp^{-1}_1: & \Psi\mapsto T_1(\Psi) &\hspace{4mm} \exp^{-1}_1(\psi)=\frac{\theta}{\sin(\theta)}(\psi-\cos(\theta))
,~\theta=d(1,\psi) \hbox{ and } \psi\in\Psi,
\end{aligned}$$
where $d(1,\psi) = \cos^{-1}\left(\int_0^1 \psi(t) dt \right)$. 
Mapping $\gamma$ onto the tangent space $T_1(\Psi)$ (summarized in Figure \ref{fig:mappings}) gives a convenient representation of $\gamma$ in the parametric vector space, and thus allows for a straight-forward prior specification to be placed on $T_1(\Psi)$ (discussed further in \ref{sec:modelspec}).
Intuitively, this linearization is akin to mapping the points on the top half of the sphere to the tangent plane at the the north pole. 
Moreover, because $T_1(\Psi)$ is the space of square integrable functions with mean zero, one can parameterize this space of functions with a basis representation.

Finally, through all of these transformations, the SRVF of $f_2 \circ \gamma$, denoted as $\mathcal{G}(g)$ can be written as
$$\begin{aligned}
\mathcal{G}(g(t))&:=(q_2,\gamma)(t)= q_2(\gamma(t))\sqrt{|\dot{\gamma(t)}|}&\\
&=q_2\bigg(\int_0^t\psi^2(s)ds\bigg)\psi(t)=q_2\bigg(\int_0^t\exp^2(g)(s)ds\bigg)\exp(g)(t),&&\\
\end{aligned}$$
where $g$ lives on the tangent space of $\Psi^+$, which is again linear.

\section{Pairwise registration in $\mathbb{R}^1$}\label{sec:pairwise}

First, we will detail our proposed approach in the simple case of pairwise alignment of functions $y_1$ and $y_2$ on $\mathbb{R}^1$.
The functions are observed at a set of discretized time points $[t] = \{t_1,\dots,t_N\}$. 
In this section, we will fully specify our proposed Bayesian hierarchical framework for this pairwise alignment setting.

Our  Bayesian approach looks for optimal warping functions that warp one function to another using the $\mathbb{L}^2$ distance of the SRVF representations of the two functions, as defined in the previous section. Furthermore, by mapping the space of warping functions to the linear tangent space $T_1(\Psi)$ on the infinite dimensional half sphere we can perform the Bayesian inference over a parametrized, linear space.

\subsection{Model specification}\label{sec:modelspec}

Let $y_1$ and $y_2$ be noisy observations of $f_1, f_2 \in \mathbb{F}$, respectively.
Then, at the first level of our hierarchical model we have,

\textit{Level 1.}
$$y_1([t])=f_1([t])+\epsilon_1([t]),\ \epsilon_1([t]) \sim MVN(0_N,\sigma_1^2 I_N)$$
and
$$y_2([t])=f_2([t])+\epsilon_2([t]),\ \epsilon_2([t]) \sim MVN(0_N,\sigma_2^2 I_N),$$
where the observed noise processes $\epsilon_1, \epsilon_2$ are are assumed to follow a Gaussian white noise distribution with variances $\sigma_1^2$ and $\sigma_2^2$, respectfully, for each point in time. 
Here we assume no temporal correlation between noise parameters.

Using the SRVF representation described in Section \ref{sec:review}, we aim to align $q_2(\gamma([t]))=q(f_2(\gamma([t]))$ to $q_1([t])=q(f_1([t]))$.
Using the transformations from the previous section, we can equate $q_2(\gamma([t]))$ to $\mathcal{G}(g([t]))$ where $g$ is the projection of the square root density of $\gamma$ onto the tangent space of the infinite half sphere and so we model $q_1([t])-\mathcal{G}(g([t]))$ using a zero-mean multivariate Gaussian distribution following \cite{lu2017}. It should be noted that $q_2(\gamma([t])) = q_2(\int_0^{[t]}\psi(s)\,ds)\psi([t])$, where $\int_0^{[t]}\psi(s)\,ds$ denotes the $N$-dimensional vector \\ $\{\int_0^{t_1}\psi(s)\,ds,\dots,\int_0^{t_N}\psi(s)\,ds\}$.
Thus, at the second level we have,

\textit{Level 2.}
$$q_1([t])-\mathcal{G}(g([t])) \sim MVN(0_N,\sigma^2 I_{N}),$$

with negative log-likelihood

\begin{equation}\label{eq:phi}
\Phi(g) := -\log p(q_1,q_2; g)\propto\bigg(\frac{1}{\sigma^2}\bigg)^{N}\exp\bigg\{-\frac{1}{2\sigma^2}\sum_{i=1}^{N}\big(q_1(t_i)-(q_2,\gamma)(t_i)\big)\bigg\}.
\end{equation}

Following the approach of \cite{lu2017}, we will assign a zero-mean Gaussian process prior to the sampled tangent space $T_1(\Psi),$
$$g \sim GP(0,\mathcal{C}),$$ 
with $\mathcal{C}$ being a positive, self-adjoint and trace-class operator on $\mathbb{R}^1.$
Zero-mean multivariate normal priors are assigned to mean functions $f_1([t])$ and $f_2([t])$ with structured kernel functions $K_1$ and $K_2$, i.e.
$$f_1([t])\sim MVN(0_N,K_1) \hbox{ and } f_2([t]) \sim MVN(0_N,K_2).$$
In this work, we specify the squared exponential kernel function $K([t])=s^2\exp(-(d/2l)^2)$ for both $K_1$ and $K_2$ where $d$ is the computed distance matrix of $[t].$
Conditionally-conjugate inverse-gamma priors are specified for all covariance parameters $\sigma^2,\ \sigma_1^2,\ \sigma_2^2$ with the prior for $\sigma^2$ less informative than prior for $\sigma_1^2$ and $\sigma_2^2$ to help with potential identifiability issues. Hyperparameters $s_1^2, s_2^2$ and $l_1,l_2$ are assigned inverse-gamma and uniform priors  respectively.

The full hierarchical posterior can now be written as 
$$p(g,\theta | y_1,y_2) = \left[ \prod_{j=1}^2p(y_j|f_j,\sigma_j^2) \right] \exp \left[ -\Phi(g) \right]\pi(g,\theta), $$
where $\theta \doteq \{f_1,f_2,\sigma^2,\sigma_1^2,\sigma_2^2,\ s_1^2,\ s_2^2,\ l_1,\ l_2\}$, $\pi(g,\theta)$ is the prior distribution over parameters $g$ and $\theta$, and $p(y_j|f_j,\sigma_j^2)=\prod_{i=1}^Np(y_j(t_i)|f_j(t_i),\sigma_j^2)$ for $j=1,2$ is a product of univariate Gaussian distributions.

\subsection{MCMC sampling}

A Metropolis within Gibbs sampler is used to sample from the complete posterior distribution $p(g,\theta|y_1,y_2).$ 
For practical implementation, functions $f_1, f_2$ are discretized and we specify a basis representation for $g$, i.e., $g=B\mathbf{v}$ where $\bf{v}$ and $B$ are the set of $n_v$ basis coefficients and matrix of basis functions, respectively. The reason for this specification is as follows. 

Recall that $g$ lives on $T_1(\Psi)$, the tangent space to the top half of the infinite-dimensional unit sphere at the point $\psi \in \Psi$. For simplicity, we take $\psi$ to be the identity element so that $T_1$ becomes the space of all $\mathbb{L}^2$ integrable functions on $[0,1]$ that have zero mean with respect to unit weight, i.e., the uniform density on the unit interval. Even though $g$ is still infinite-dimensional we can sample it using the  Karhunen-Lo\`{e}ve expansion where we specify the covariance operator $C$ via its eigenpairs $(b_k,\lambda_k^2)$, where the set $\{b_k(\cdot)\}$ forms an orthonormal basis for the functions space $T_1(\Psi)$. We therefore sample independent variates $v_k\sim\mathcal{N}(0,\lambda_k^2)$ and set
$$ g = \sum_{k=1}^M v_k b_k, $$
where $M$ is the number of basis functions. This is similar to the method proposed in \cite{lu2017}.
We can then use a Fourier series type representation to parameterize $C$ in terms of a finite collection of basis coefficients, $\mathbf{v}$, and discretized orthogonal basis vectors represented by columns of $B$. 
The eigenvalues, $\lambda_k$, are set to $\sigma_g^2/k_2$, where $\sigma_g$ is assumed to be known or provided.
The choice of orthogonal basis functions are plentiful, but this paper uses the Fourier series and Legendre polynomial expansions. Both sets form a complete basis over the space of $\mathbb{L}^2$ integrable functions on a compact interval. 
Moreover, for the application at hand, both basis functions (other than the constant) are orthogonal to the unit weight; thus making it easy to satisfy the tangent space property. 
The Fourier basis is typically used for periodic functions and so convergence can be hindered if the function values are different at the domain end points. 
Legendre polynomials do not have this limitation, but they are not as widely known. 
The smoother the function to be approximated is, the faster the decay of the basis coefficients.
This means we need fewer basis coefficients to represent the underlying function. 
In fact if the function is infinitely differentiable, the decay of the basis coefficients can be spectral, i.e., exponentially decaying. This applies to both Fourier and polynomial based expansions (see \cite{hesthaven2007}). 
The basis specification for $C$ thus allows us to more efficiently explore the posterior space $g|y_1,y_2,\theta$.

To update $g,$ we wish to sample from the conditional posterior $p(g|y_1,y_2,\theta, B)$ $\propto \exp\big(-\Phi(g)\big)$.
To account for the complex geometry of the target space which ${g|\cdot}$ lies on and to allow for our sampler, we will exploit Hamiltonian dynamics.
Specifically we use the $\infty$-HMC algorithm described in \cite{beskos2017} to sample ${g|\cdot}$ within our MCMC sampling procedure. $\infty$-HMC is a modification of the standard Hamiltonian Monte Carlo (HMC algorithm developed to be well-defined on the infinite-dimensional Hilbert Space and thus is ideal for sampling from $T_1(\Psi)$.
HMC is a natural choice for more complex geometries, as it allows for efficient sampling along posterior contours. 
As \cite{beskos2017} describes, HMC is an improvement over the preconditioned Crank-Nicholson method. 
Moreover, as we will show, in cases where the target density is multi-modal and separated by low probability regions, a parallel chain implementation with random starting points is needed to fully explore the state space of ${g|\cdot}$ (details given in Section \ref{sec:multalign}).

The algorithm, detailed in Algorithm \ref{alg:infhmc} utilizes an auxiliary variable.
As suggested in \cite{beskos2011} and implemented in \cite{beskos2017}, we choose to specify the auxiliary variable $v,$ interpreted as the \textit{velocity} of $g$.
To implement Algorithm \ref{alg:infhmc}, and to update $g$, it is necessary to define the location-specific preconditioner matrix $\mathcal{K}(g)$ as the covariance of a local Gaussian approximation $N(m(g),\mathcal{K}(g))$ to the posterior. 
We define this pre-conditioner through its inverse:

$$\mathcal{K}(g)^{-1} = \mathcal{C}^{-1} + \beta \mathcal{H}(g),$$
where  $\mathcal{H}(g)$ is chosen as the Gauss-Newton Hessian (GNH), i.e.
$$\mathcal{H}(g) = \langle \bigtriangledown \mathcal{G} (g),\ \mathcal{C}^{-1}\bigtriangledown \mathcal{G} (g)\rangle.$$
We can then calculate the natural gradient $\eta$:
$$\eta (g)= -\mathcal{K}(g) \big[ \bigtriangledown \Phi (g)-\beta \mathcal{H} (g) g\big], $$
to calculate the Hamiltonian flow, $\Xi^t.$
As the exact analytic expression of $\Xi^t$ is often not available, $\infty$-HMC defines the flows $\Xi_1^t,\ \Xi_2^t$ of a split Hamiltonian system to numerically approximate $\Xi^t:$
$$\Xi_1^t(g,v)=\big(g,v+\frac{t}{2}\eta(g)\big)\ \hbox{ and }\  \Xi_2^t(g,v) = \big(g\cos t+v\sin t,-g\sin t+v\cos t\big).$$
The leapfrog map $\Psi_h(g,v): (u_0, v_0) \mapsto (g_h,v_h)$ is the composition of three sub-steps:
$$\Psi_h(g,v)=\Xi_1^{h/2} \circ \Xi_2^h \circ \Xi_1^{h/2},$$
and the exact Hamiltonian flow $\Xi^T$ is then approximated by
$$\Psi_h^{I}(g,v)=\Psi_h^{\lfloor T/h \rfloor},$$ a concatenation of $I=\lfloor T/h \rfloor$ Verlet steps.
$\infty$-HMC requires user chosen time-step $h$ and leap-frog step $T.$ $\bigtriangledown\Phi(g)$ is the directional derivative of $\Phi$, defined in \eqref{eq:phi}, with respect to $g$ which is provided in the \ref{sec:deriv}.

Compared to the \textit{Z-mixture pCN algorithm} \citep{cotter2013} used in \cite{lu2017} to update $\bf{v}$, we find that using $\infty$-HMC algorithm to update $g$ is significantly more efficient. 
We favor $\infty$-HMC over Z-mixture pCN because we are sampling from an infinite dimensional space. \cite{beskos2017} demonstrates that random walk-like algorithms, like the Z-mixture pCN, are not eﬀicient and can break down on these high dimensional spaces. 
Additionally, the gradient information  $\infty$-HMC relies upon helps inform the sampler and has been shown to increase overall MCMC sampling performance for hierarchical models, especially when sampling on infinite dimensional spaces \citep{beskos2017, BDA}. 
For broader application and sampling eﬀiciency of the prior on the space of diffeomorphisms we chose to implement $\infty$-HMC for these reasons.
    
We found that the Z-mixture pCN algorithm, even when implemented via parallel chains with random starts, was not able to capture the expected bimodal posterior in our simulated data example. 
The posterior was biased towards one mode only as shown in the posterior samples and achieved higher SSE. Figures demonstrating these comparisons are given in \ref{sec:mcmcdiag}. 
The additional tuning parameters, specifically the leap frog and time step sizes ($T$ and $h$ in Algorithm~\ref{alg:infhmc} respectively), available using $\infty$-HMC allowed for the posterior samples to more easily jump between the multiple possible target modes.

\begin{algorithm}
\caption{Sampling Algorithm for $g$ via $\infty$-HMC (\cite{beskos2011})}\label{alg:infhmc}

\begin{enumerate}
\item Given current $g$, propose $g'= P_g\{\Psi_h^{I}(g,v))\}$  where $P_g\{\cdot\}$ is the projection of $(g',v')=\Psi_h^{I}(g,v)$ onto the $g$ argument and $v$ (velocity) is an auxiliary variable sampled from $N(0,\mathcal{C}).$\\

\item Accept $g'$ with probability $1 \wedge \exp\{-\Delta H(g,v)\},$
$$\Delta H(g,v)\equiv H(g',v')-H(g,v),\ \ H(g,v)=\Phi(g)+\frac{1}{2}\langle g,\mathcal{C}^{-1}g\rangle+\frac{1}{2}\langle v,\mathcal{C}v\rangle.$$

\end{enumerate}

\end{algorithm}

Mean functions $f_1, f_2 \in \mathbb{R}^1$ are sampled from their respective conditional posteriors
$$p(f_k|y_k,\theta)\propto p(y_k|f_k,\sigma_k^2)p(q_k([t])-\mathcal{G}(g([t]))|y_1,y_2,\theta)\pi(f_k),\ k=1,2~,$$
via Metropolis-Hastings with specified proposals 
$$f_k'([t]) \sim MVN(f_k([t]),\Sigma_k^*),\ k=1,2,$$ 
and squared exponential correlation structures specified for $\Sigma_1^*$ and $\Sigma_2^*.$

Variance parameters $\sigma^2,\ \sigma_1^2,\ \sigma_2^2$ and hyperparameters $s_1^2$ and $s_2^2$ can be Gibbs sampled from their respective inverse-gamma posteriors. Hyperparameters $l_1$ and $l_2$ can be sampled via Metropolis-Hastings with Gaussian proposal distributions and acceptance ratios:

$$\alpha_{l_k}=\frac{p(f_k|y_k,l'_k)\pi(l_k)}{p(f_k|y_k,l_k)\pi(l'_k)},\ k=1,2.$$

\subsection{MCMC for multiple optimal alignments}\label{sec:multalign}

Due to the high efficiency of the $\infty$-HMC algorithm, a single posterior chain generated using $\infty$-HMC would not be able to jump between modes and thus would be unable to capture multi-modal distributions. 
As highlighted by \cite{nishimura2017}, if the target distribution is multi-modal with modes separated by regions of low probability density; it is nearly impossible for any HMC algorithm to transition between modes due to the conservation of energy property of Hamiltonian dynamics.
To obtain a multi-modal posterior distribution for $g|y_1,y_2,\theta,$ we run $K$ parallel chains with initial values for the basis coefficients of $g$ chosen randomly from a standard Normal distribution.

The idea is to run separate chains at enough randomly generated starting points to land in the vicinity of the multiple modes; so that we don't have to rely on the HMC algorithm to cross the low-probability regions. Of course, we have no idea where those modes are in general, so it is important to explore the initial starting space. 
In this case the starting space is the space of coefficients, $\mathbf{v}$, of the linear expansion of $g$. Furthermore, under the assumption that each chain reaches their target sampling density after a thorough burn-in period, we can treat each chain as providing independent samples; and thus simply pool them together. For example, if each of the ten chains produce 10k samples, we can pool them together to obtain 100k samples. 
Alternatively, we can think of this approach as a type of parallel MCMC with a constant temperature parameter and no mixing.

It is important to highlight that there could be more than one best possible solution. Having a posterior that can capture the total variability of the warping function sample all possible solutions, rather than simple pick one as a local optima, is an advantage of our method. However in practice, it may be necessary to choose between multiple possible solutions. This determination is application dependent; a good metric is to choose the best warping function that produces the smallest amplitude distance $||q_1 - (q_2,\gamma))||^2$ between the two SRVFs $q_1$ and $q_2$. Another suggestion is one sees see more of the posterior favoring one alignment over the other and this solution could be chosen.


\section{Examples}\label{sec:examples}

To demonstrate our proposed method, we will compare our flexible Bayesian approach with the state-of-the-art Dynamic Programming (DP) method of \cite{FSDA}. In short, the DP algorithm solves the optimal warping function by proposing successively better piecewise linear pathways on a unit square grid; where better is defined in terms of the Fisher-Rao metric.

First, we will assess our method's performance on simulated noisy data for which multiple optimal alignments exist. Next, we will apply our method to two real datasets: a SONAR dataset and an iPhone movement dataset. The SONAR data is notably noisy and thus difficult to register. The iPhone data is less noisy but has multiple optimal alignments that previous methods discussed in Section \ref{sec:intro} do not account for.

Before we move on to the examples, we need to address the issue of computing means or averages in non-Euclidean geometries, e.g., the positive orthant of infinite-dimensional unit sphere. For problems of this type, we typically use the Karcher or Fr\'{e}chet mean. 
This is obtained by minimizing the average squared distance between all pairs of sample points, $x_1,\dots,x_N \in \Psi$ i.e.
$$m = \text{arg} \min_{p \in \Psi} \sum_{i=1}^N d^2(p,x_i), $$
where $d$ is the metric or geodesic distance measure on $\Psi$, the infinite-dimensional positive orthant of the unit sphere, given by
$$d(\psi_1,\psi_2) = \cos^{-1}\left(\int_0^1 \psi_1(s)\psi_2(s) ds \right).$$
Note that in the case of Euclidean geometry, the Karcher mean is simply the average. The computation of the Karcher median is analogous. We use the method presented in \cite{XieKurtek:2016}, which is an extension of Algorithm 2 in \cite{tucker2013}. Now, in the multiple alignment case, our posterior will be multi-modal. In order to find the centroid of these individual modes, we do the following. We begin by computing all the pairwise distances under the metric $d$, between all posterior samples. We then use a clustering algorithm to separate the samples into their respective clusters. Then, for each cluster the Karcher mean or median can be computed as in the unimodal case.

\subsection{Simulated data} \label{sec:simex}
\begin{figure}[H]
\centering
\includegraphics[width=0.48\textwidth]{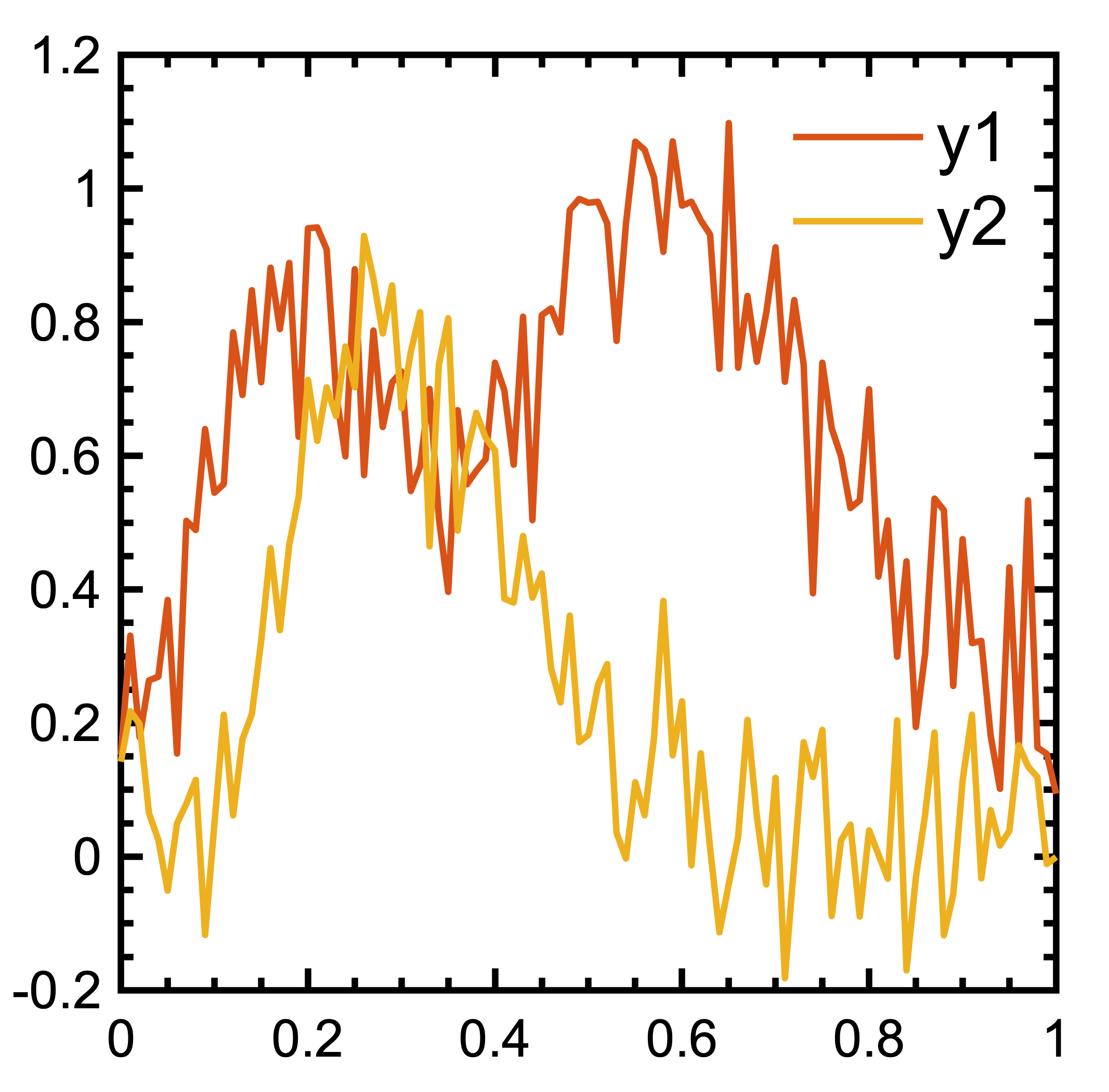}
\includegraphics[width=0.48\textwidth]{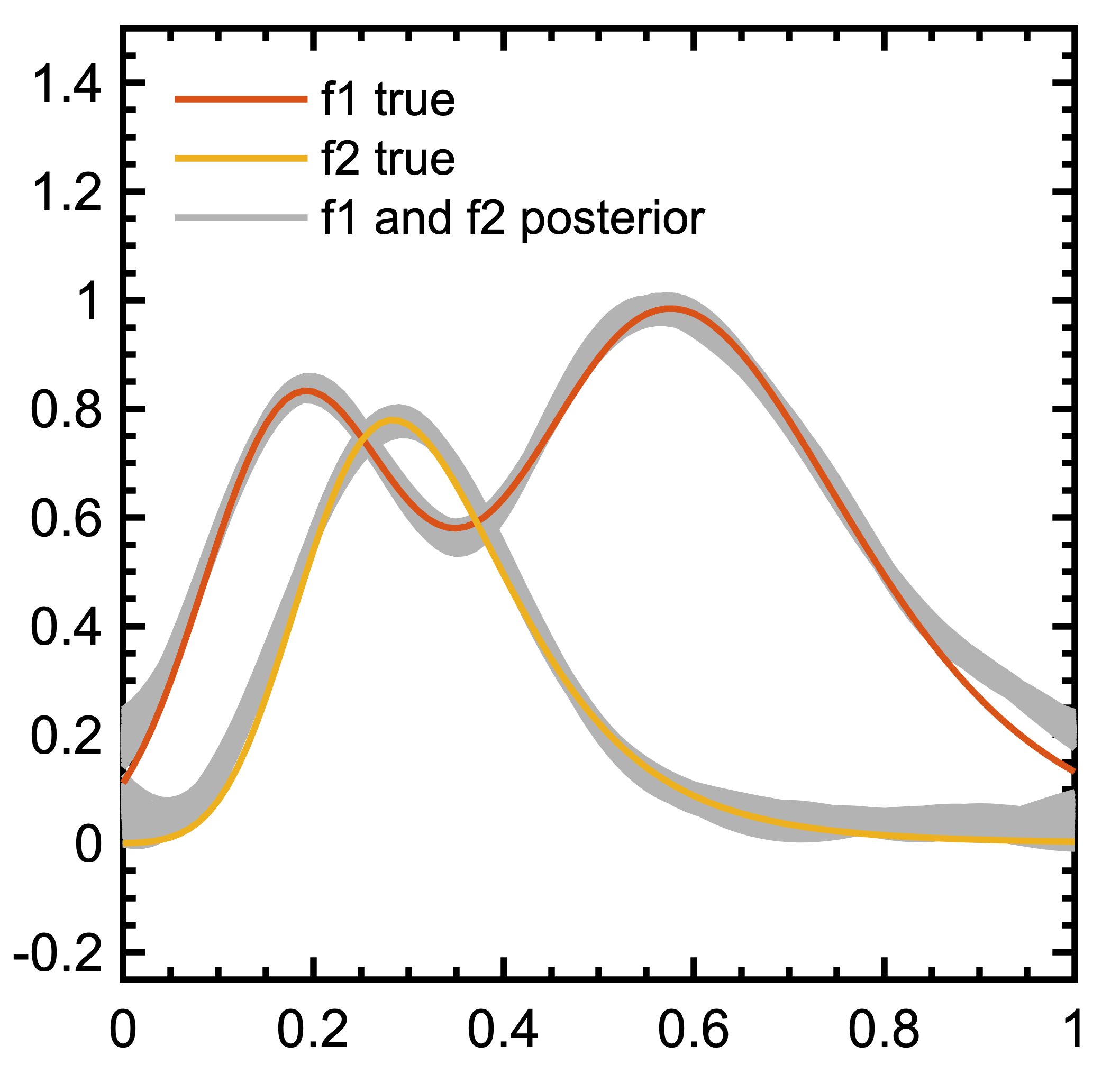}
\caption{Simulations $y_1$ and $y_2$ (left) and their respective posteriors $f_1|y_1$ and $f_2|y_2$ (right) plotted against the true functions.}
\label{fig:simf1f2}
\end{figure}

Consider the misaligned and noisy observations $y_1$ and $y_2$ shown in Figure \ref{fig:simf1f2} simulated by adding Gaussian white noise with $\sigma_1^2=\sigma_2^2=0.001$ to the true mean functions $f_1$ and $f_2$. Without loss of generality, we wish to align $y_2$ to $y_1.$

\begin{figure}[H]
\centering
\includegraphics[width=0.48\textwidth]{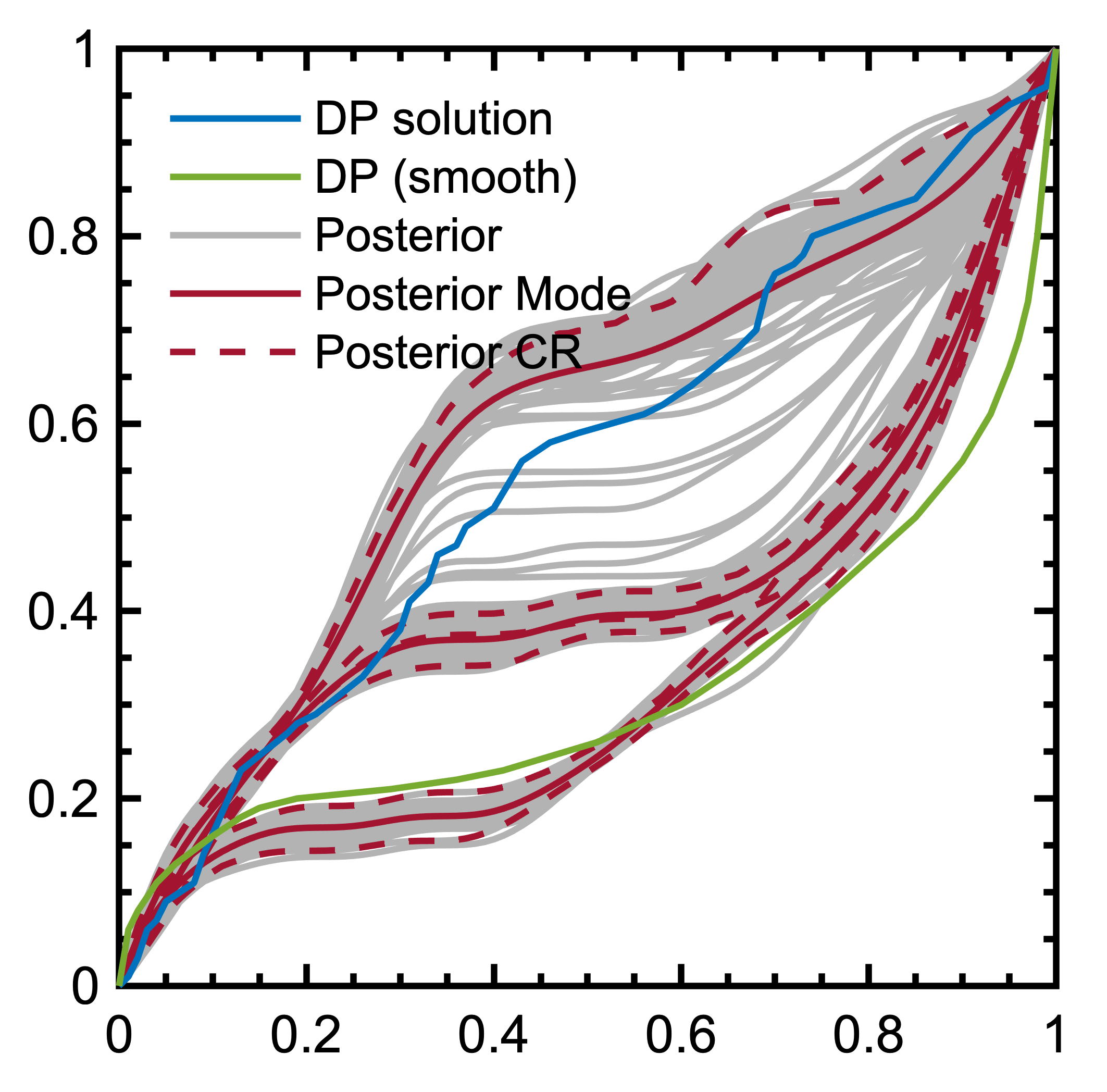}
\includegraphics[width=0.48\textwidth]{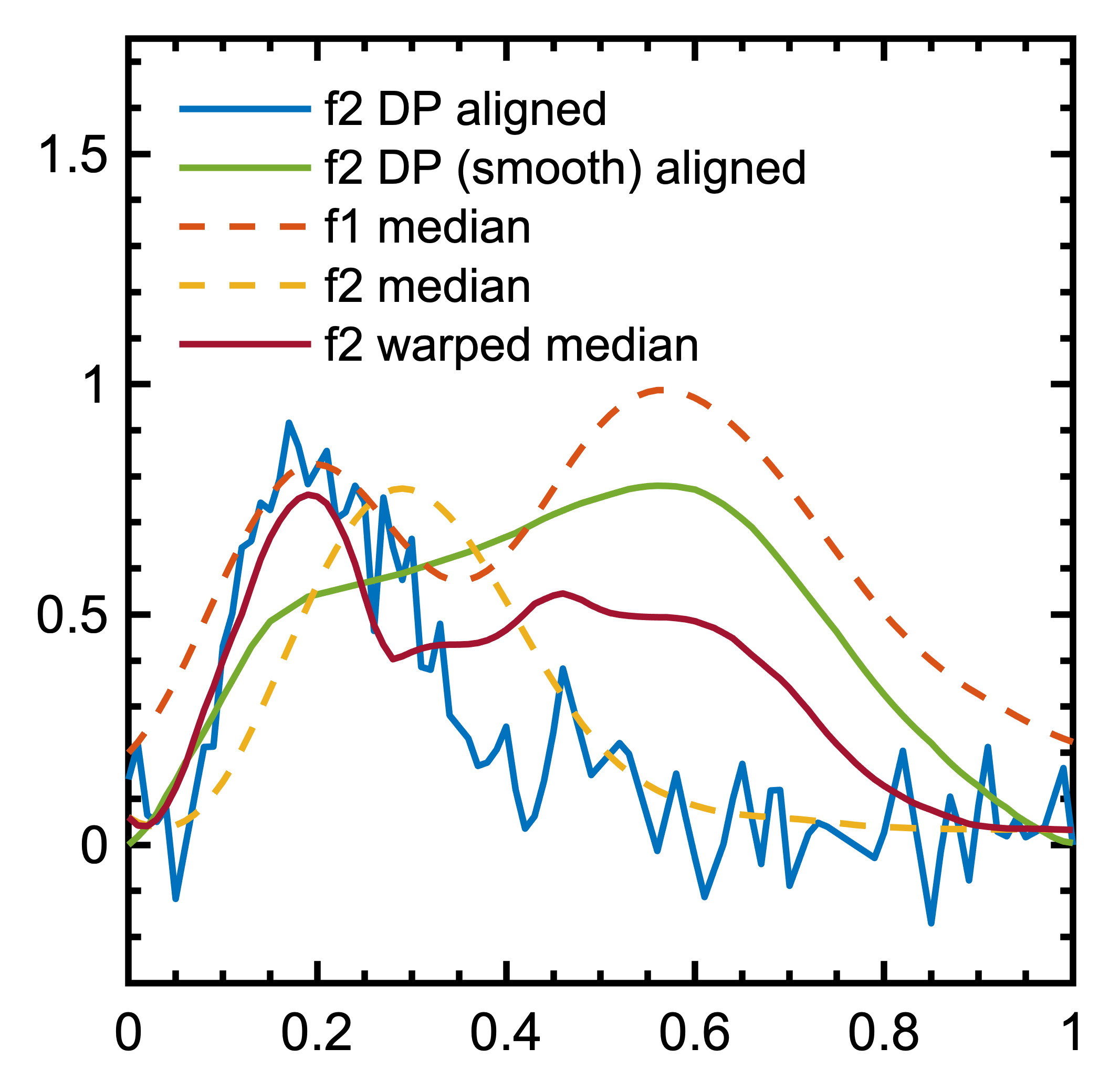}
\caption{Comparison of the posterior distribution $\gamma|y_1,y_2$ to the DP solution $\gamma_{DP}$ applied to the simulated data, plotting the two identified modes and respective 95\% credible regions of the posterior $\gamma|y_1,y_2$ (left). Comparison of the DP warping of $y_2$ and smoothed $y_2$ to the median of the warping $f_2 \circ \gamma|y_1,y_2$ (right).}
\label{fig:simwarping}
\end{figure}

The right panel in Figure \ref{fig:simf1f2} shows the estimated posteriors for $f_1$ and $f_2$ given our Bayesian approach. 
Figure \ref{fig:simwarping} compares the dynamic programming (DP) solution with our Bayesian approach. 
On the left, we see the optimal warping function for $y_2$ in blue, found by DP ($\gamma_{DP}$), which reflects some of the noise in the data and only identifies one possible warping function. 
The corresponding warped solution $y_2 \circ \gamma_{DP}$ is also shown in blue on the right.
Also the warped solution of $\hat{f}_2 \circ \gamma_{DP}$ is shown in green, where $\hat{f}_2$ smoothed version of $y_2$ and DP is performed on a smoothed versions of $y_1$ and $y_2$. 
The smoothing in this case was performed using Gaussian process regression.
In our Bayesian approach, we ran 8 MCMC chains each with 20,000 iterations, a burn-in of 5,000 iterations without thinning, for a total of 4,000 effective samples. 
The proposal parameters were chosen such that acceptance rates for each chain were between 0.2 and 0.4 (see Figure \ref{fig:simex_chains} for MCMC chains). 
Furthermore, we used $n_v=10$ Fourier basis functions to approximate the function $g$ on the tangent space $T_1(\Psi)$ resulting in a dimensionality of 20 (two for each basis function). The left figure of Figure \ref{fig:simwarping} compares the posterior samples of $\gamma|y_1,y_2$, with credible intervals and posterior mode to the DP solution. The right figure of Figure \ref{fig:simwarping} compares the posterior median warped solution to the DP warped solutions. From this we can see that the Bayesian approach gives us three possible solutions: (1) $f_2$ is aligned to  the leftmost peak of $f_1$, (2) $f_2$ is aligned to the rightmost peak of $f_1$, and (3) $f_2$ is aligned to both peaks of $f_1$. Note that the DP method only gives us one out of the possible three solutions.

\begin{figure}[H]
\centering
\includegraphics[width=0.48\textwidth]{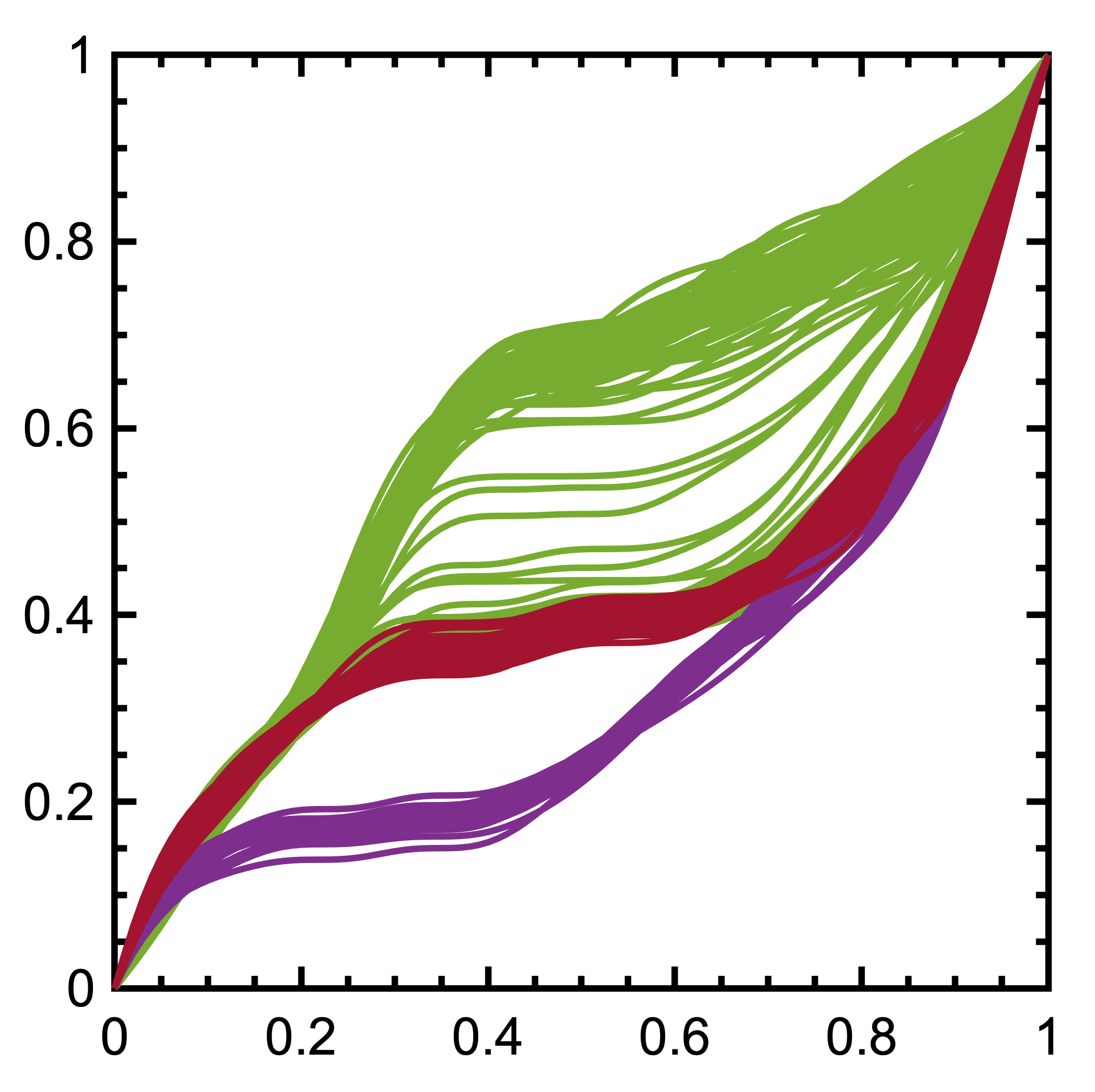}
\includegraphics[width=0.49\textwidth]{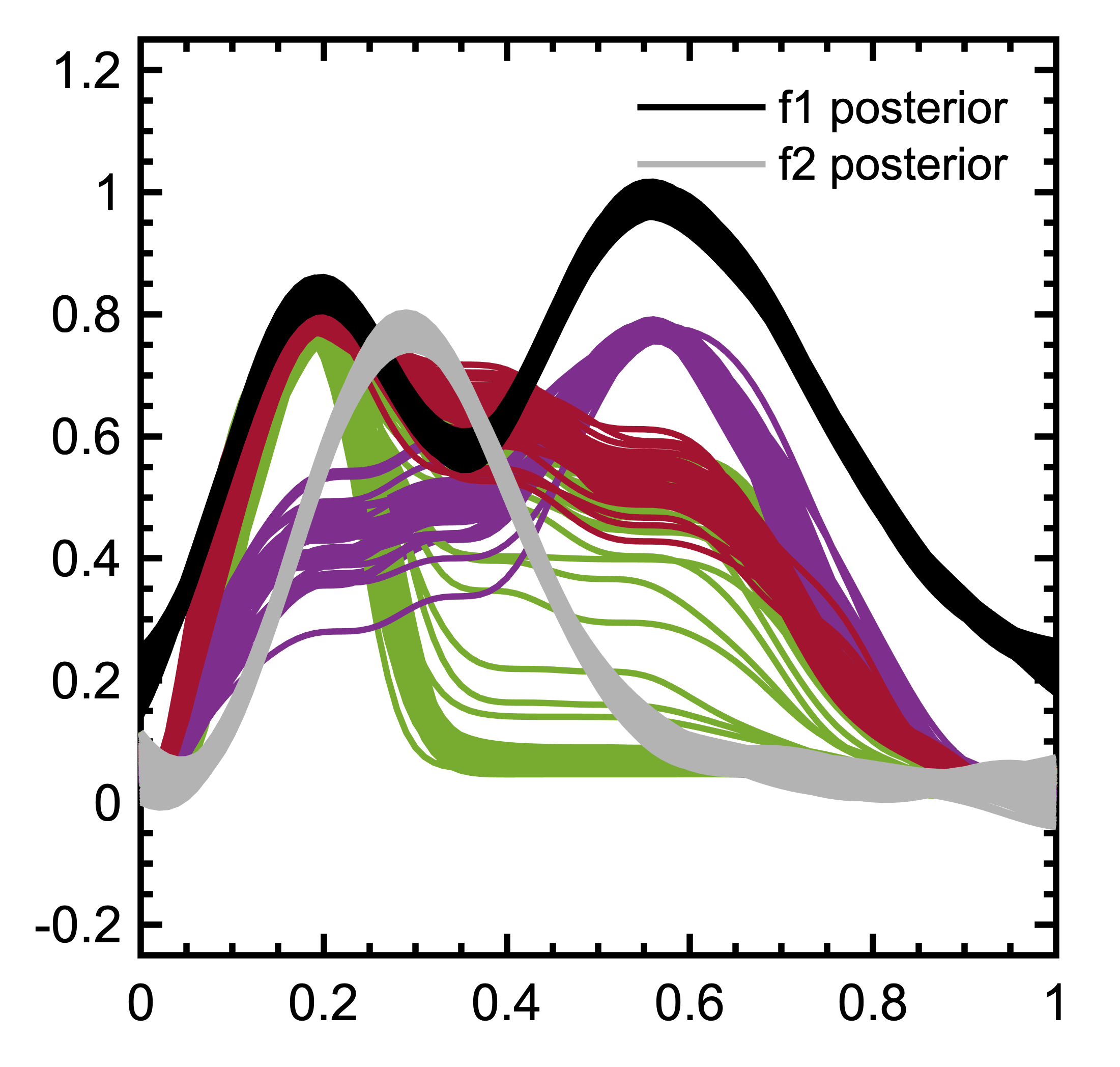}
\caption{$\gamma|y_1,y_2$ (left) and $f_2 \circ \gamma|y_1,y_2$ (right) from the 8 different MCMC chains, colored by the 3 posterior modes found.}
\label{fig:simcolored}
\end{figure}

Further exploration of the Bayesian approach is shown in Figure \ref{fig:simcolored}. Here, we break down the posteriors $\gamma|y_1,y_2$ and $f_2 \circ \gamma|y_1,y_2$ from the 8 different MCMC chains, colored by the 3 posterior modes found. It is clear that 3 unique alignments were captured in the posteriors of $\gamma|y_1,y_2$ and $(f_2\circ \gamma)|y_1,y_2$ among the 8 different MCMC chains; identified by three distinct colors. The left shows the three posterior modes and respective 95\% credible regions, characterizing the uncertainty of each possible warping and the uncertainty of $(f_2\circ \gamma)|y_1,y_2$.
This is exhibited by the width of the color-coded samples on the right.
To give a statement on computational cost our implementation of the MCMC algorithm is done in MATLAB; and one chain for $3.2e4$ iterations takes 24 seconds on a 8 Core Mac Pro with 32GB of RAM. 
The dynamic programming algorithm is computationally $O(N^2)$, where $N$ is the dimension of the grid in the search space for the minimal energy, and for this simulated example $N=101$. 
The code provided the \verb+fdasrvf+ MATLAB package for DP is written in C and runs in 0.2 seconds.

\begin{figure}[H]
  \centering
  \includegraphics[width=\textwidth]{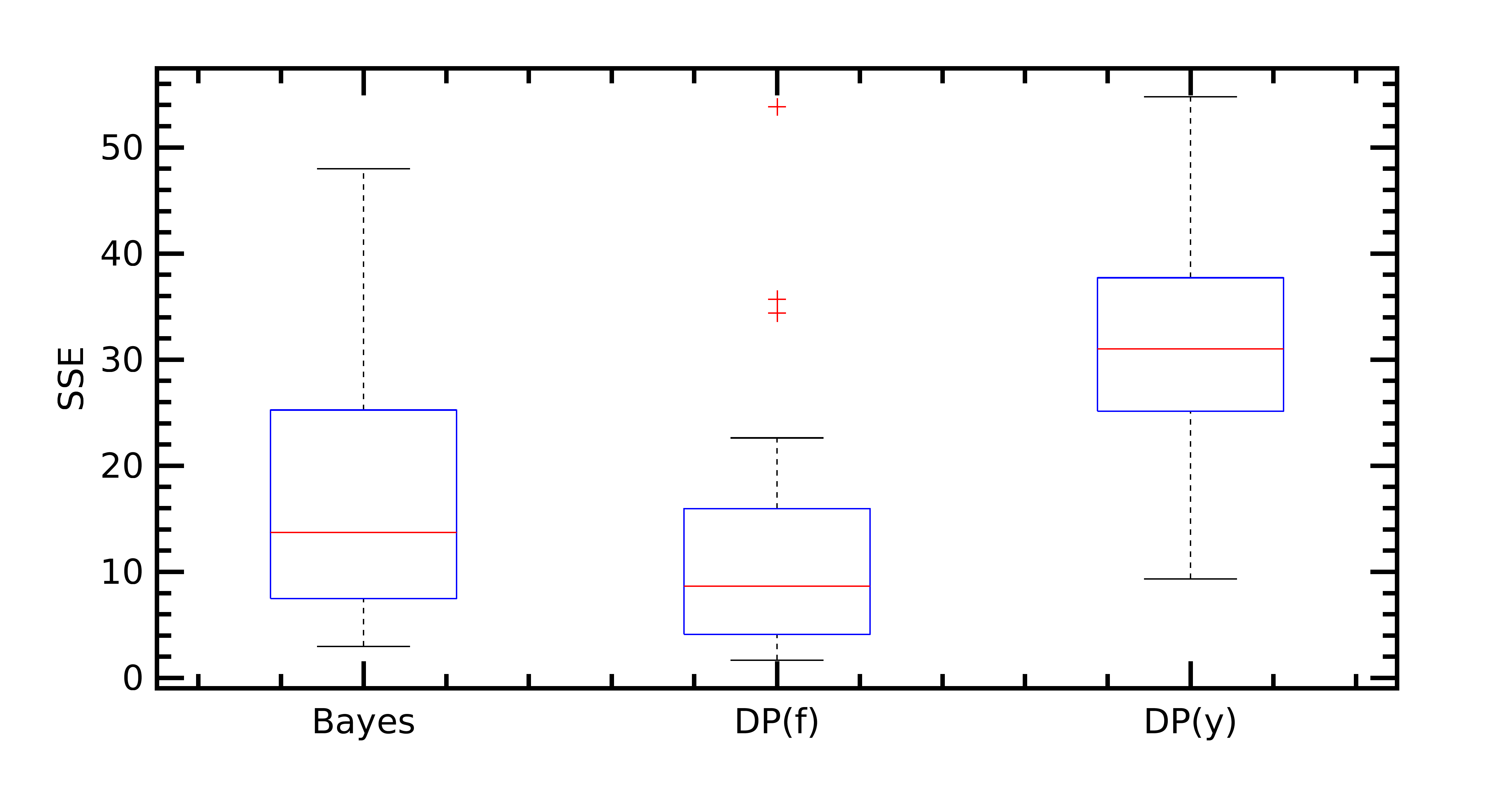}
  \caption{Boxplot of the sum-of-squared errors for the Bayesian method compared to DP on the noisy data $y_j,~j=1,2$ (DP(y)) and DP on the smoothed noisy data (DP(f)).}
  \label{fig:multsim_SSE}
\end{figure}

We executed this simulation for 25 replicates and computed the point-wise sum-of-squared error (SSE) for the median warping from our method and the warping function from DP performed using the raw data, $y_j,~j=1,2$; also from performing DP on the smoothed version of $y_j$. 
Figure~\ref{fig:multsim_SSE} presents a box plot for the SSE for all three methods. 
The SSE of the Bayesian method is similar to performing DP on the smoothed data. 
It should be noted that there is a larger spread in the inter-quartile region of the Bayesian method, and this is due to the fact that it is capturing multiple possible solutions that have different SSE values. 
Both of the DP solutions only find one solution, and this uncertainty (e.g., more than one registration) is not characterized in these methods. 

\begin{figure}[H]
  \centering
  \begin{minipage}{\textwidth}
    \centering
    \includegraphics[width=.48\textwidth]{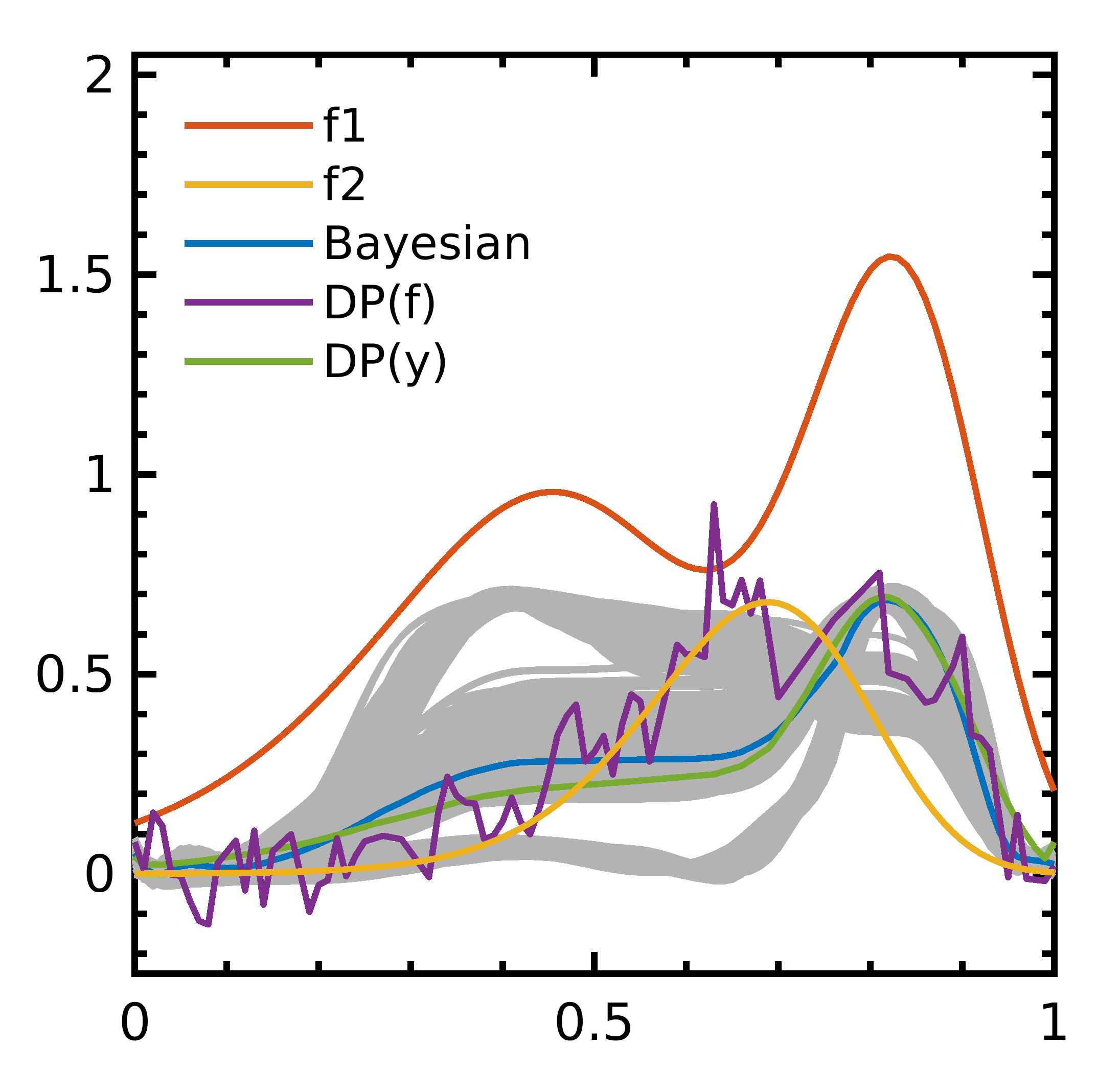}\quad
    \includegraphics[width=.48\textwidth]{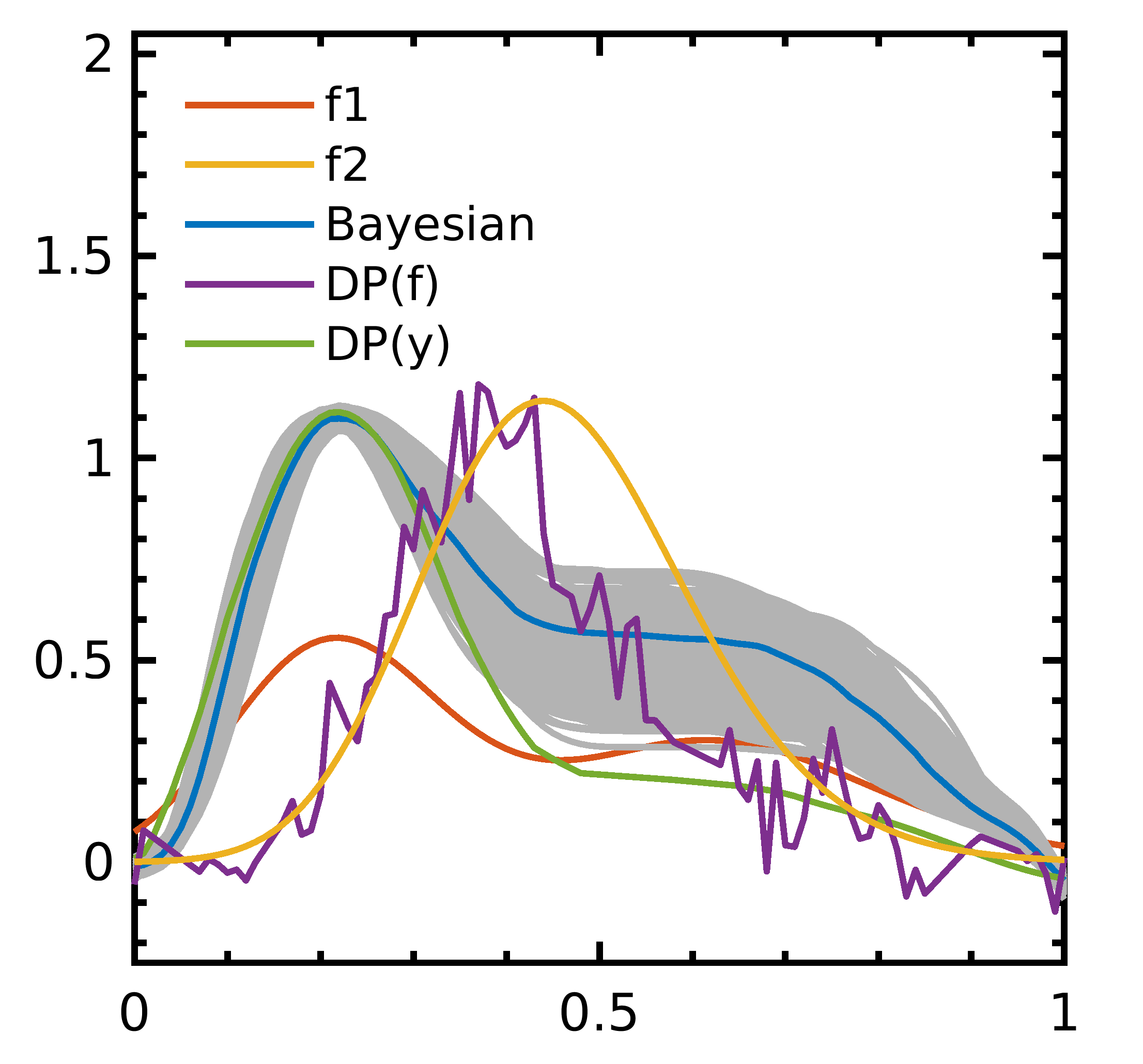}\\
    \includegraphics[width=.48\textwidth]{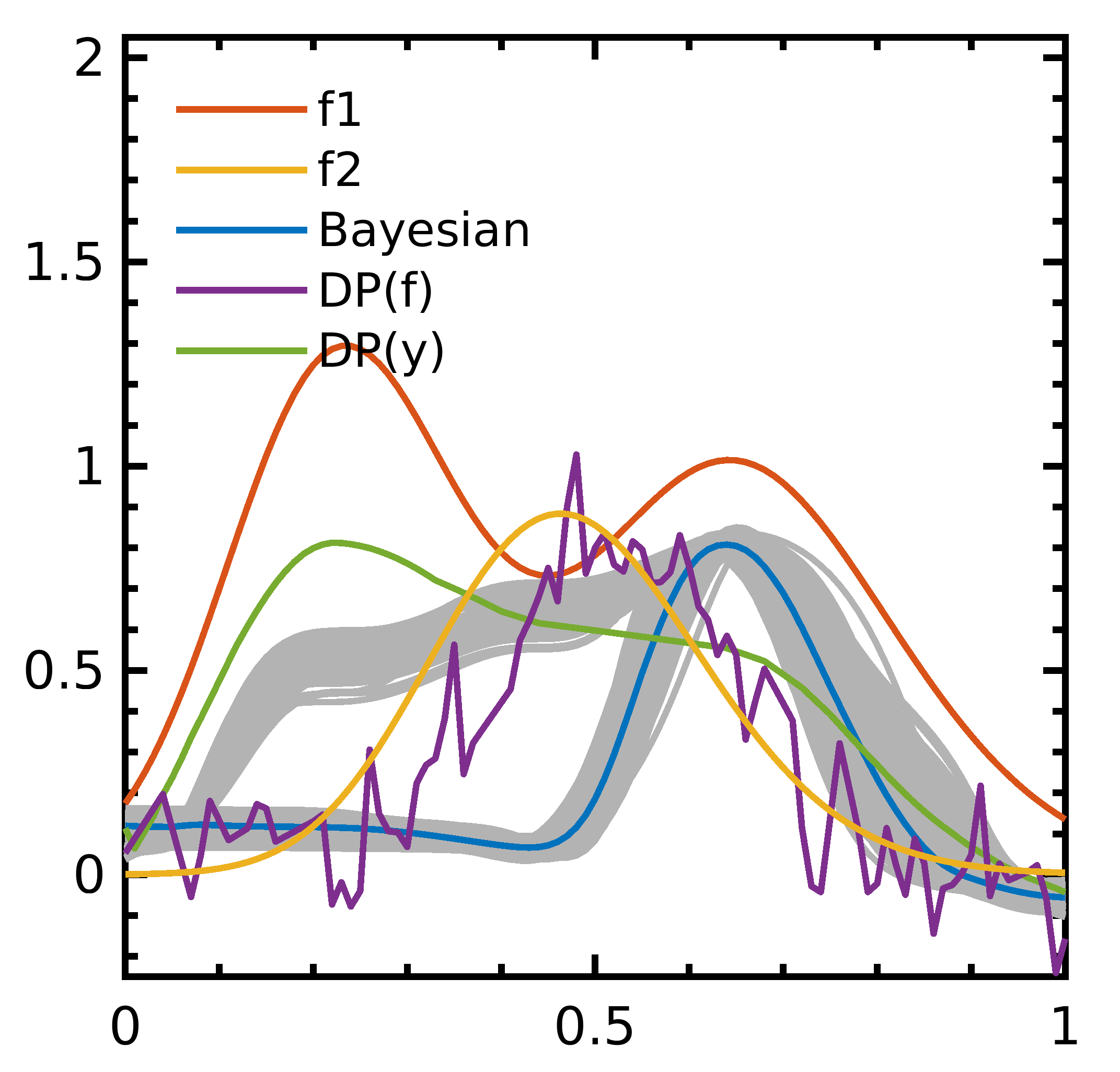}\quad
    \includegraphics[width=.48\textwidth]{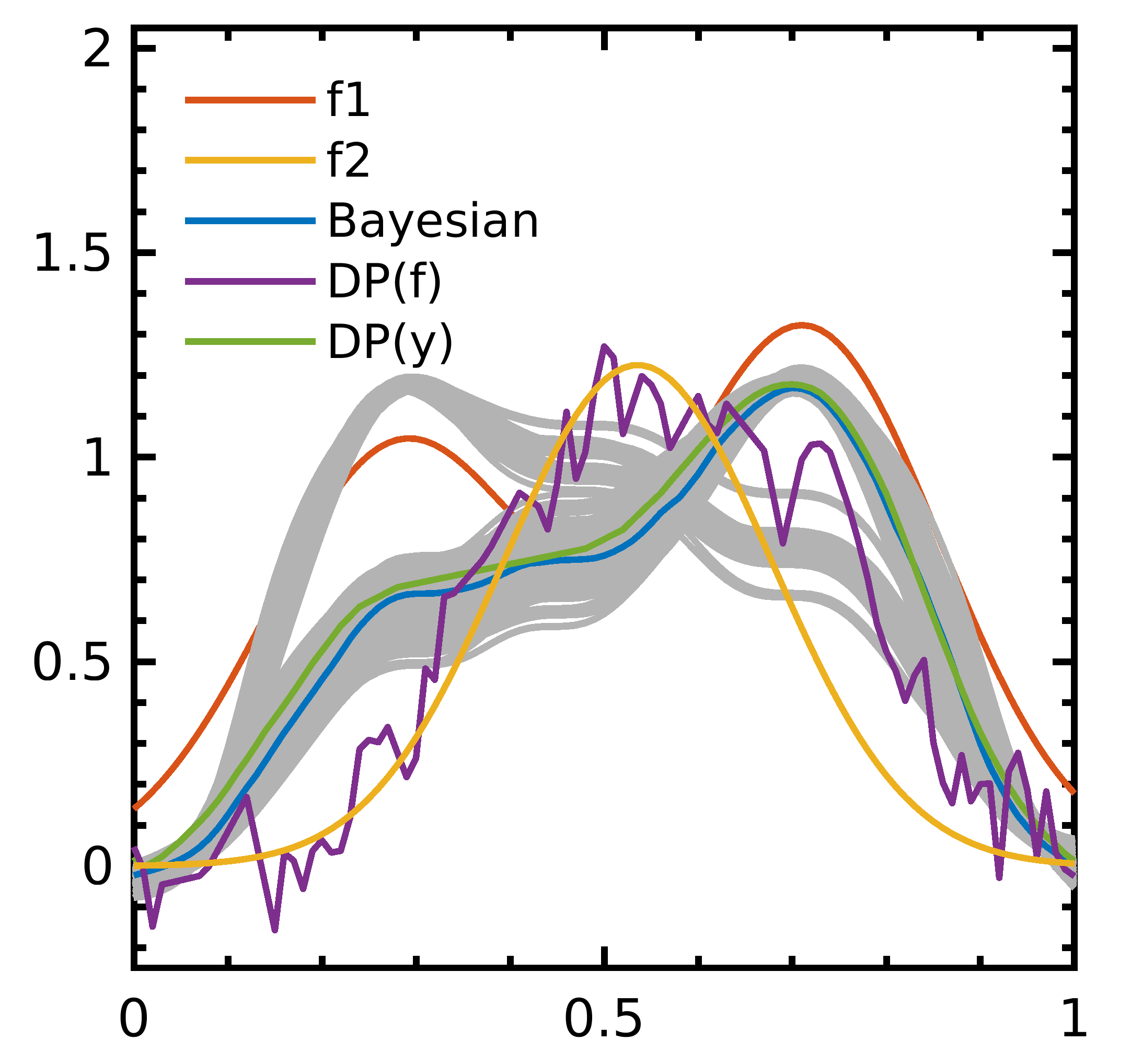}
  \end{minipage}
  \caption{Four replicates from the simulated data showing the Bayesian alignment with the posterior in gray, also shown is the DP solution on both the noisy and smoothed noisy data.}
  \label{fig:multsim_ex}
\end{figure}

Figure~\ref{fig:multsim_ex} presents four replicates from the simulated data with the median from the Bayesian posterior along with the posterior samples.
The median of the posterior of $y_1$ and $y_2$ is shown respectively as $f_1$ and $f_2$. 
Additionally, the DP solutions for the noisy data (green) and the smoothed noisy data (blue). 
As expected the median of the Bayesian solution and the DP on the smoothed noisy data match nicely.
However, the benefit of the Bayesian solution is shown in the posterior, where the multiple possible registrations are captured (with associated uncertainties) as the example as $f_1$ has two peaks and $f_2$ has one peak.

\subsection{SONAR data} \label{sec:sonar}
Next, we apply our alignment approach to naturally noisy SONAR data collected at the Naval Surface Warfare Center Panama City Division's (NSWC PCD) test pond. 
For a description of the pond and measurement setup, the reader is referred to \cite{art:kargl}. 
In summary, the idea is to use the acoustic signals, generated from SONAR data, to discriminate and classify underwater ``targets'', e.g., unexploded ordnance. More precisely, acoustic signals were generated from the raw SONAR data to construct ``target strength" as a function of frequency and aspect angle. 
Due to the relatively small separation distances between the targets in the measurement setup and orientation, and uncertainty in the placement, the scattered fields from the targets overlap leading to misaligned signatures.
This could lead to erroneous identification and classification of the target. 
For this example, we took two one-dimensional misaligned signatures, each with 1102 data samples, from a target that was a small notched aluminum cylinder.

In our Bayesian approach, we ran 8 MCMC chains with 4,000 iterations each and a burn-in period of 2,000 steps without thinning; for a total of 16,000 samples. 
The proposal, prior, and hyperprior parameters were chosen such that acceptance rates were between 0.2 and 0.4 (see Figure \ref{fig:sonar_chains} for MCMC chains). 
We additionally specified $n_v=4$ Fourier basis functions to approximate $g$, as opposed to 8 in the previous example. 
We observed that reducing the number of basis functions $n_v$, allowed our HMC algorithm to more easily explore the posterior space of $g$; likely by allowing the inference to focus entirely on the lower frequency basis elements. 
Lastly, due to the noisiness of this data we only sampled $20\%$ of the data, instead of the full set of 1102 data points, in order to obtain smooth estimates of $f_1$ and $f_2$ and avoid overfitting.

\begin{figure}[H]
\centering
\includegraphics[width=0.48\textwidth]{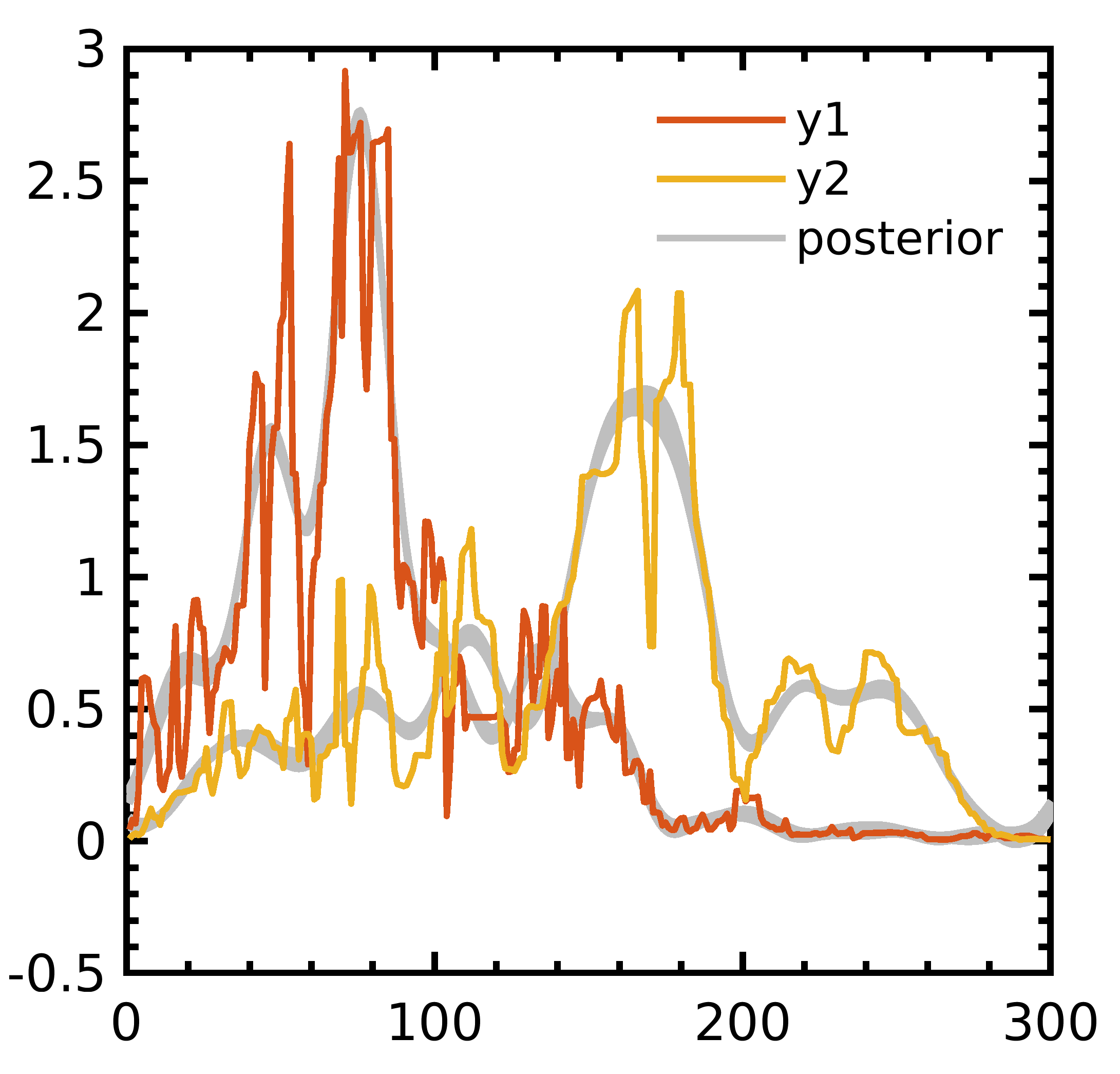}
\includegraphics[width=0.48\textwidth]{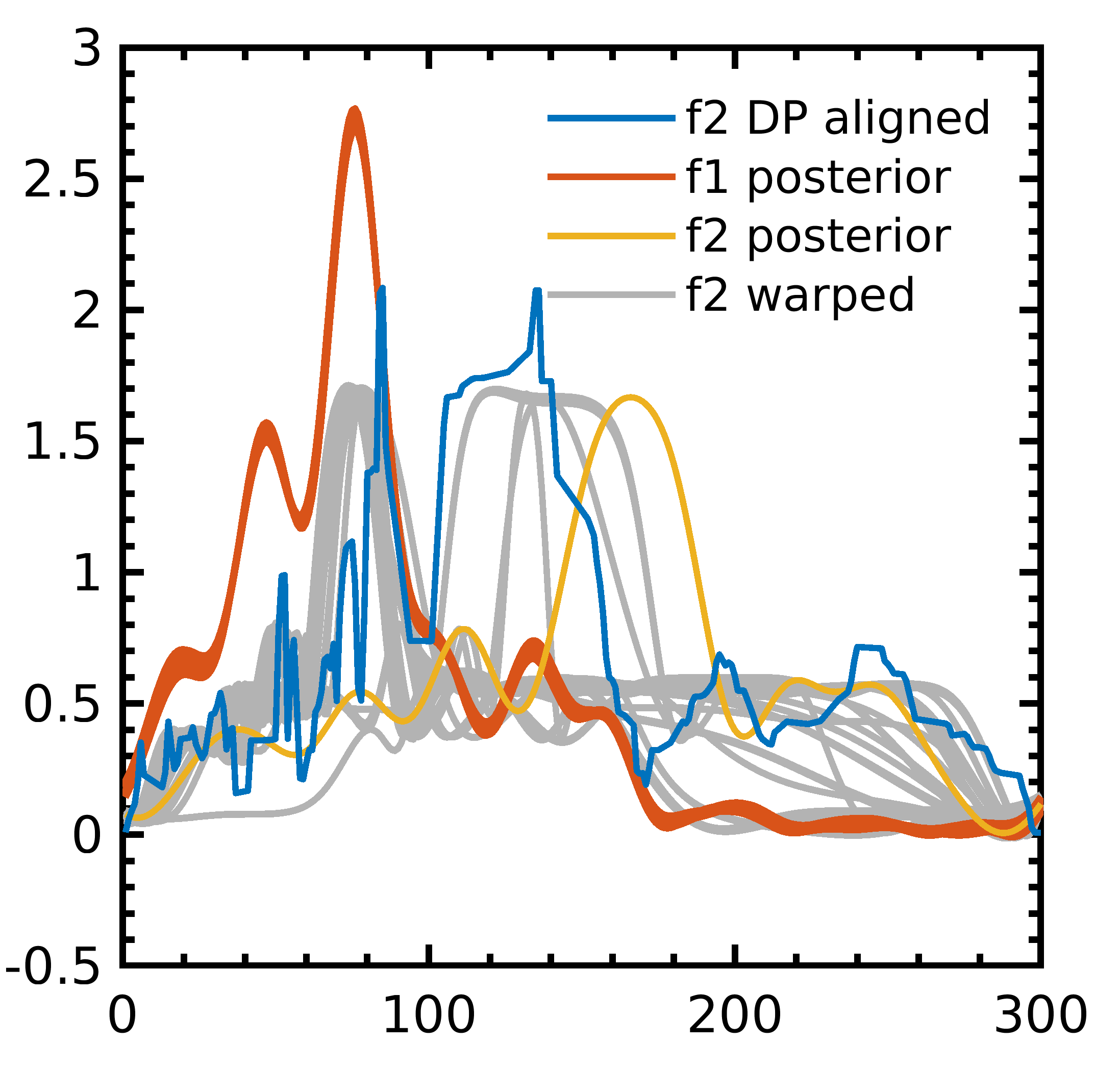}
\caption{Posteriors $f_1|y_1$ and $f_2|y_2$ shown in gray compared to the raw data $y_1$ and $y_2$ (left). The posterior of the warped function  $(f_2 \circ \gamma)|y_1,y_2$ captures more possible alignments than the single DP warped solution in blue (right).}\label{fig:sonar_f1f2post}
\end{figure}

The posteriors $f_1|y_1$ and $f_2|y_2$ of the acoustic signatures from the SONAR data are shown in Figure \ref{fig:sonar_f1f2post} on the left (gray) and right (in orange and yellow). On the right, we compare the DP aligned solution (in blue) and the Bayesian aligned solution (in gray). Note that our Bayesian approach offers multiple possible alignments, one of which is similar to the DP approach, but another which more accurately matches the maximum peaks of $f_1$ and $f_2$.

\begin{figure}[H]
\centering
\includegraphics[width=0.48\textwidth]{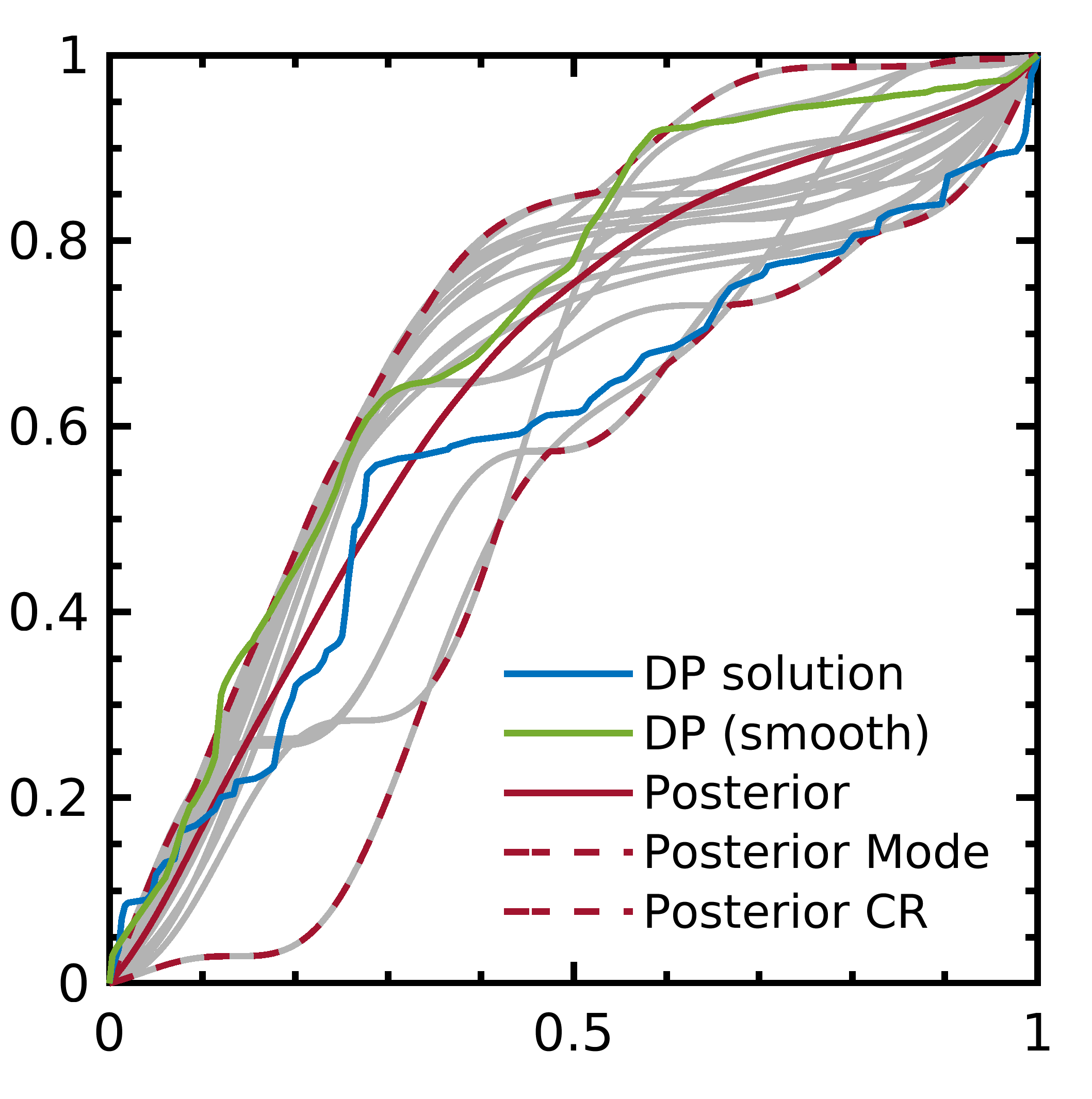}
\includegraphics[width=0.49\textwidth]{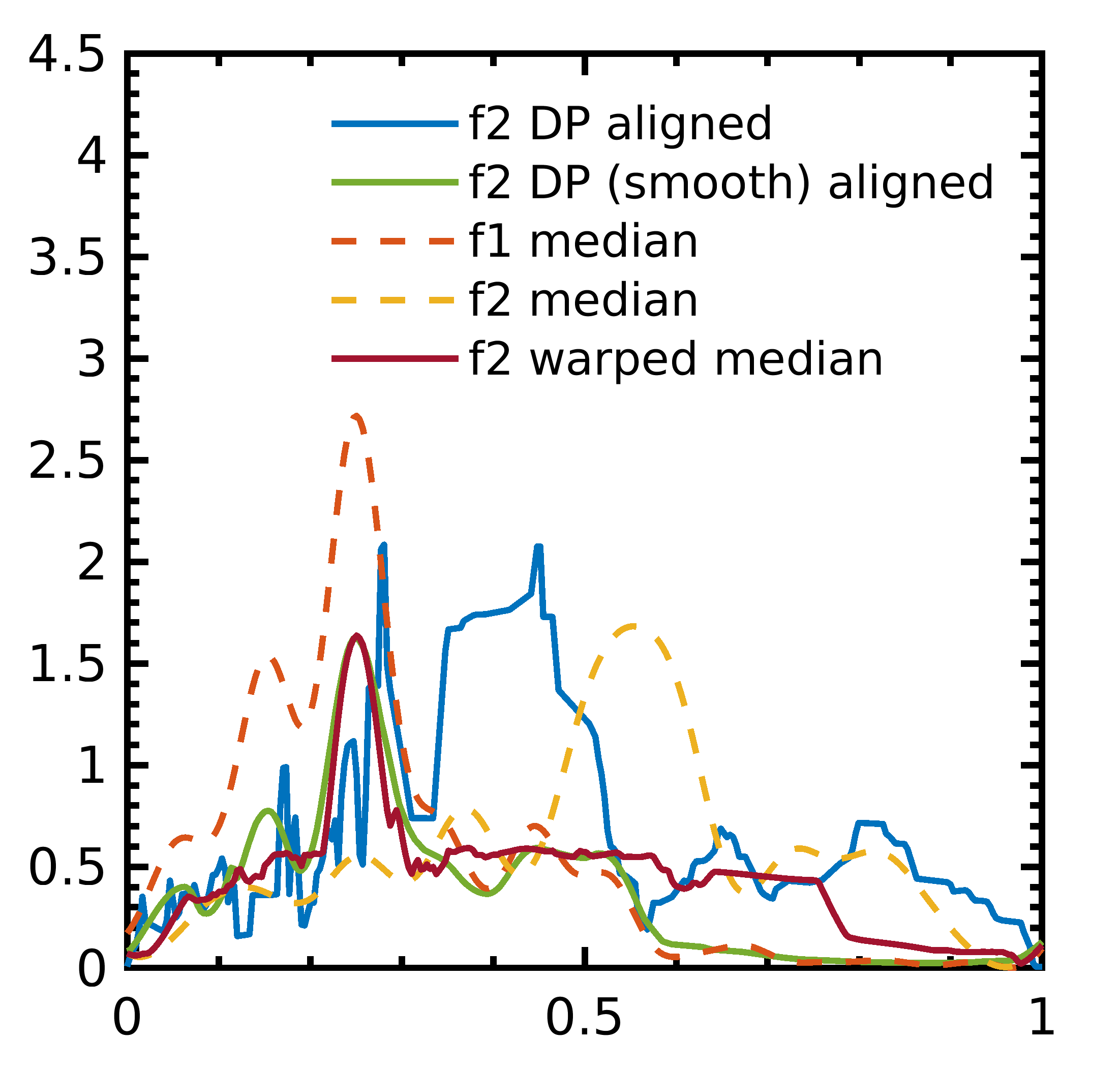}
\caption{Comparison of the posterior distribution $\gamma|y_1,y_2$ to the DP solution $\gamma_{DP}$ applied to the SONAR data, plotting the two identified modes and respective 95\% credible regions of the posterior $\gamma|y_1,y_2$ (left). Comparison of the DP warping of $y_2$ and smoothed $y_2$ to the median of the warping $f_2 \circ \gamma|y_1,y_2$ (right).}
\label{fig:sonar_gampost}
\end{figure}

Figure \ref{fig:sonar_gampost} on the right panel compares the MAP estimates to the DP solution. Figure \ref{fig:sonar_gampost} on the left compares the DP solution to the posterior distribution of $\gamma|y_1,y_2$ and shows the posterior median and $95\%$ credible region of $\gamma|y_1, y_2$. 
The credible region quantifies the uncertainty around the estimated warping function $\gamma$ and thus the warped function $(f_2 \circ \gamma)$. 
Note that the credible region obtained from the Bayesian approach encapsulates the DP solution, which itself is not able to adequately account for the noise in the data. 
Subsequently, the uncertainty can be pushed through any later analyses for any quantity of interest that may depend on the aligned or warped functions. 
For example, any statistical moment calculations performed on the aligned functions, e.g., principal component analysis, can be modified to account for the uncertainty in the warping functions.\footnote{The most straightforward way to do this would be in Monte Carlo type fashion - compute the statistics for your quantity of interest (QoI) for each posterior sample $\gamma|y_1,y_2$, resulting in a histogram for the desired QoI.}   

\subsection{iPhone data} \label{sec:iphone}
This data set consists of aerobic actions of subjects, such as biking, running, walking, etc., recorded using the Inertial Measurement Unit (IMU) on an Apple iPhone 4 smartphone; which these days is like using a cassette tape to listen to music. 
The IMU included a 3D accelerometer, gyroscope, and magnetometer. 
Each sample was taken at 60Hz, and manually trimmed to 500 samples (every 8.33s) to eliminate starting and stopping movements. 
For more information on the data set the reader is referred to \cite{mccall-reddy-shah}. 

\begin{figure}[H]
\centering
\includegraphics[width=0.48\textwidth]{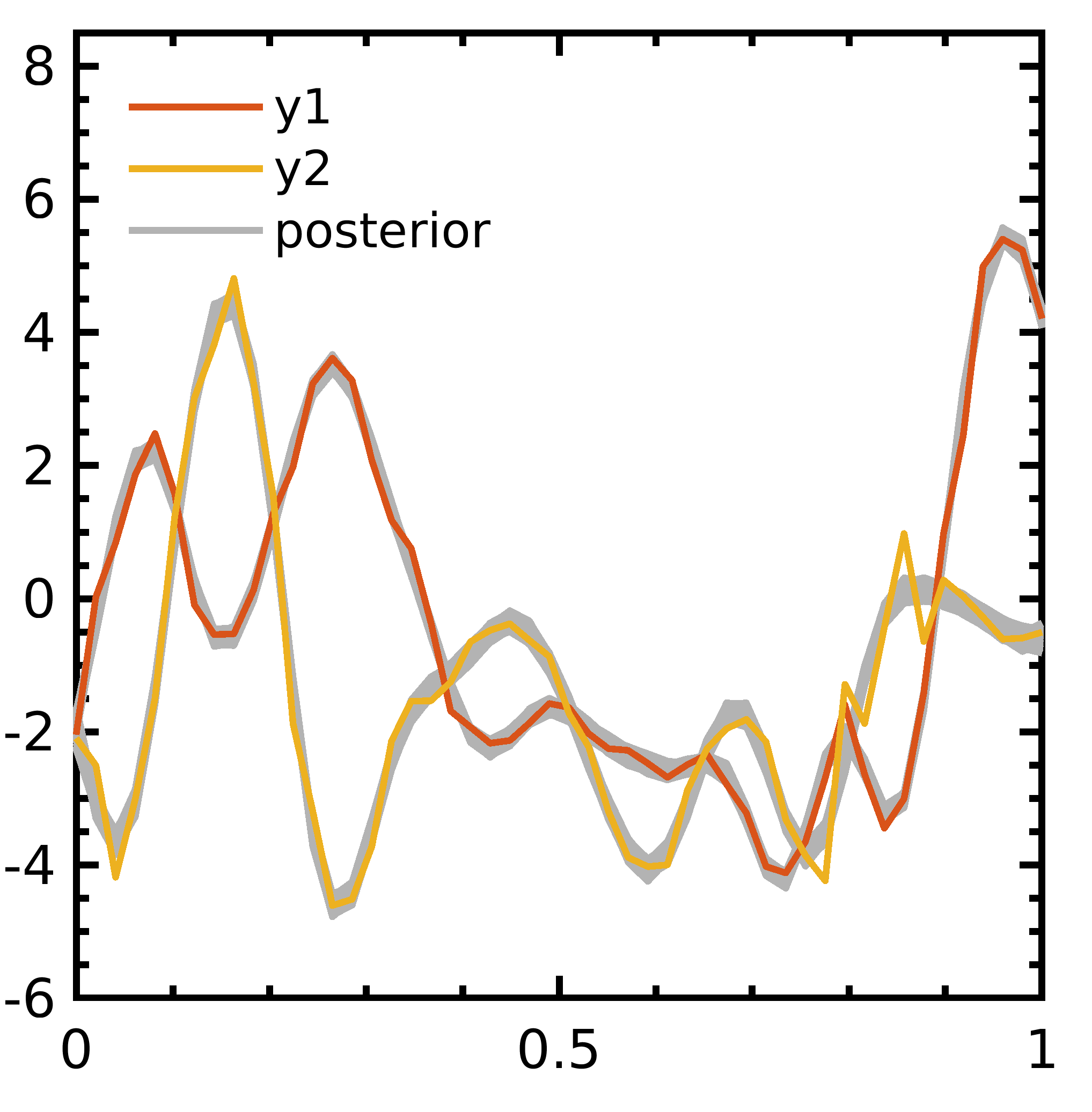}
\includegraphics[width=0.48\textwidth]{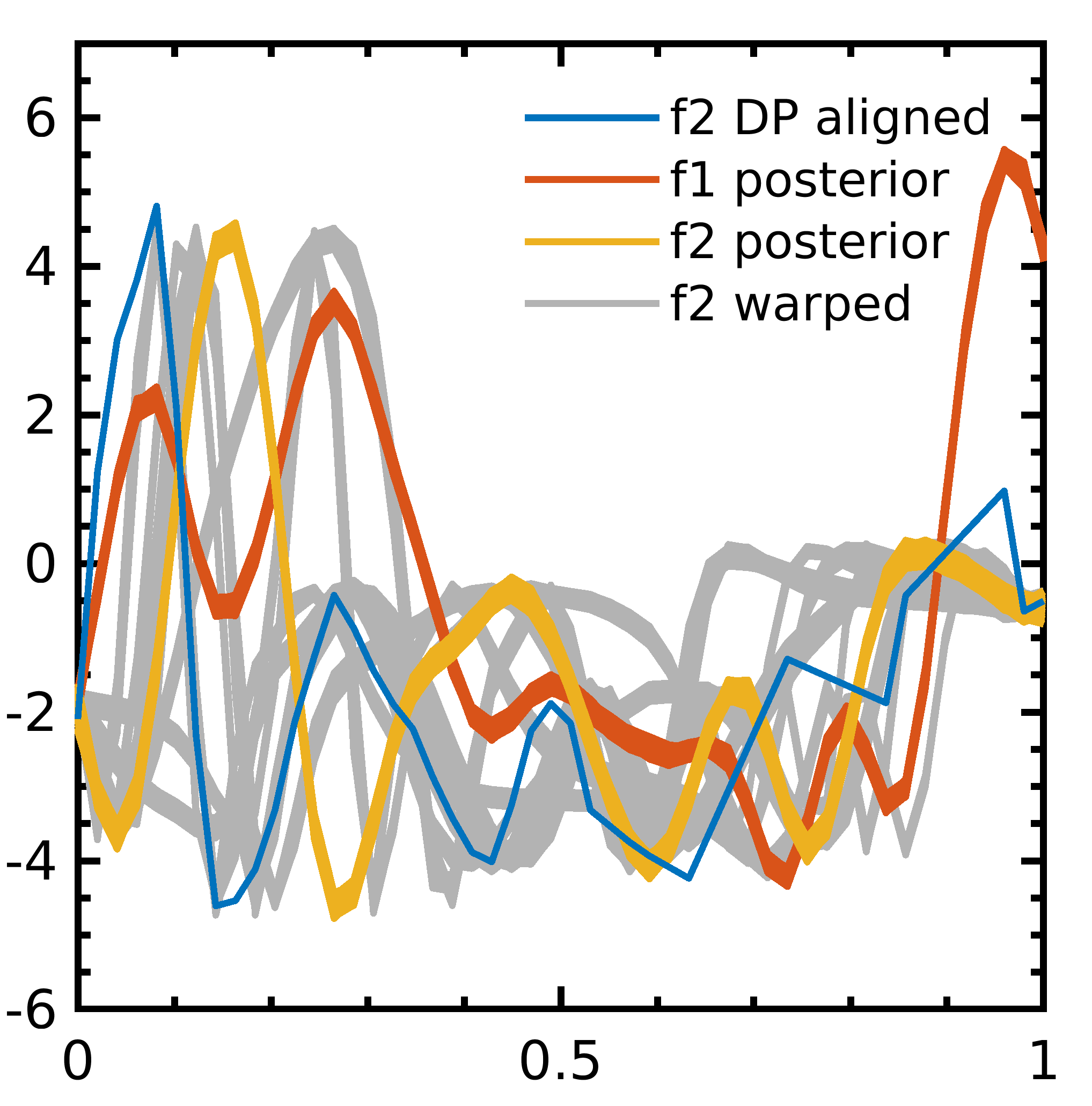}
\caption{Posteriors $f_1|y_1$ and $f_2|y_2$ shown in gray compared to the raw data $y_1$ and $y_2$ (left). The posterior of the warped function  $(f_2 \circ \gamma)|y_1,y_2$ captures a more accurate alignment than the single DP warped solution in blue (right).}
\label{fig:iphone_f1f2post}
\end{figure}

We chose to demonstrate our method on two of forty-five functional samples from the walking accelerometer data in the $x$-direction. In this context, we are specifically interested in the information contained in the separated phase and amplitude components, rather than the resulting alignment. 
For example we are only interested in the amplitude variability of the acceleration in the $x$-direction. Comparing the functions after warping would allow us to consider this variability.
The two examples we are aligning, shown in the right figure of Figure \ref{fig:iphone_f1f2post}, clearly have multiple possible alignments of $y_2$ to $y_1$ that we would like to also capture in the posterior $\gamma|y_1,y_2.$

To perform the Bayesian inference, we again ran 8 MCMC chains with 5,000 iterations each, a burn-in period of 1,000 steps, and thinning every other sample, for a total of 16,000 effective samples. 
The proposal, prior and hyperprior parameters were chosen such that acceptance rates were between 0.2 and 0.4 (see Figure \ref{fig:iphone_chains} for MCMC chains). 
We additionally specified $n_v=8$ Legendre basis functions to approximate $g.$ We specify Legendre polynomials here since the data is non-periodic in nature.

\begin{figure}[H]
\centering
\includegraphics[width=0.48\textwidth]{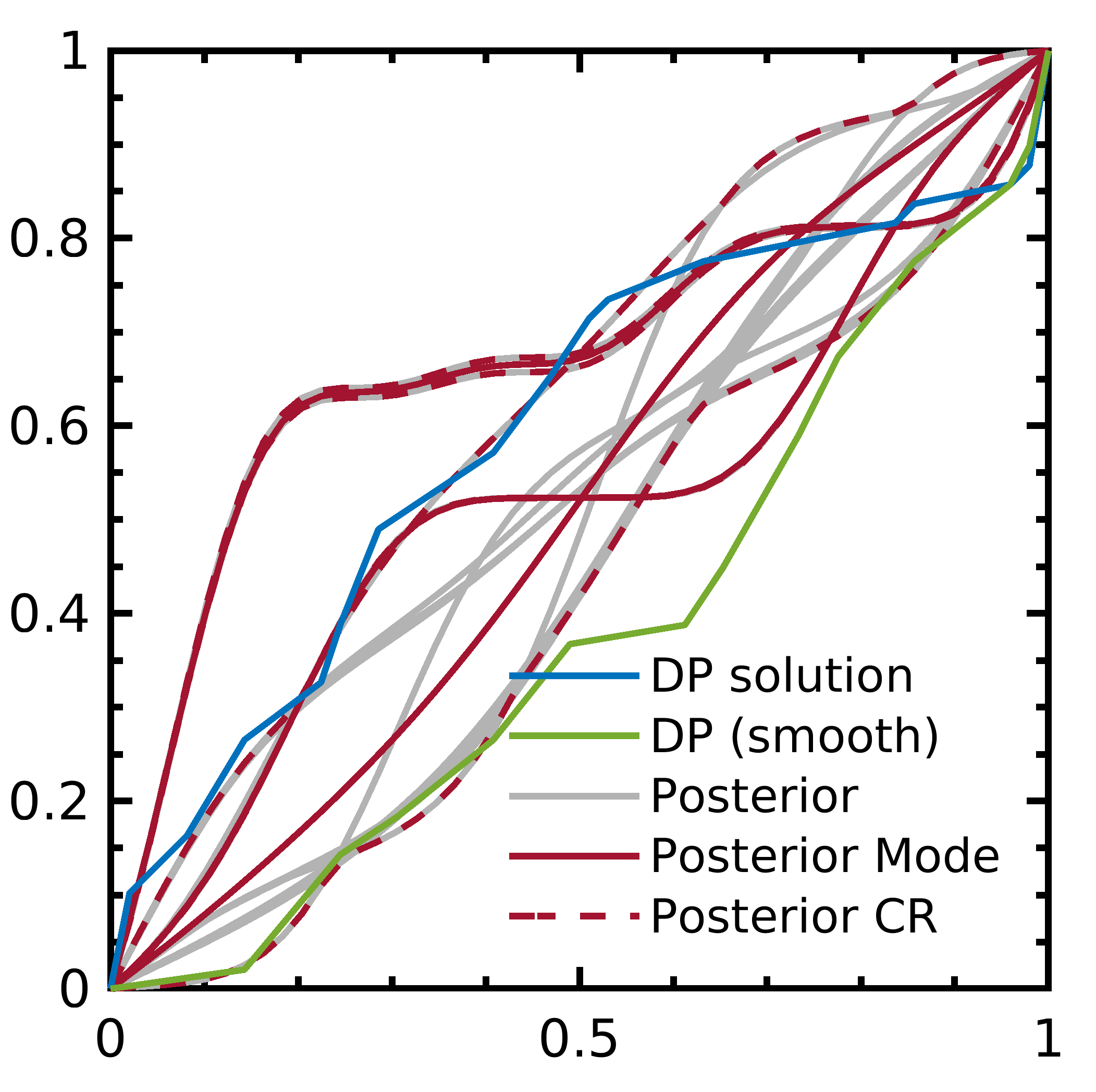}
\includegraphics[width=0.48\textwidth]{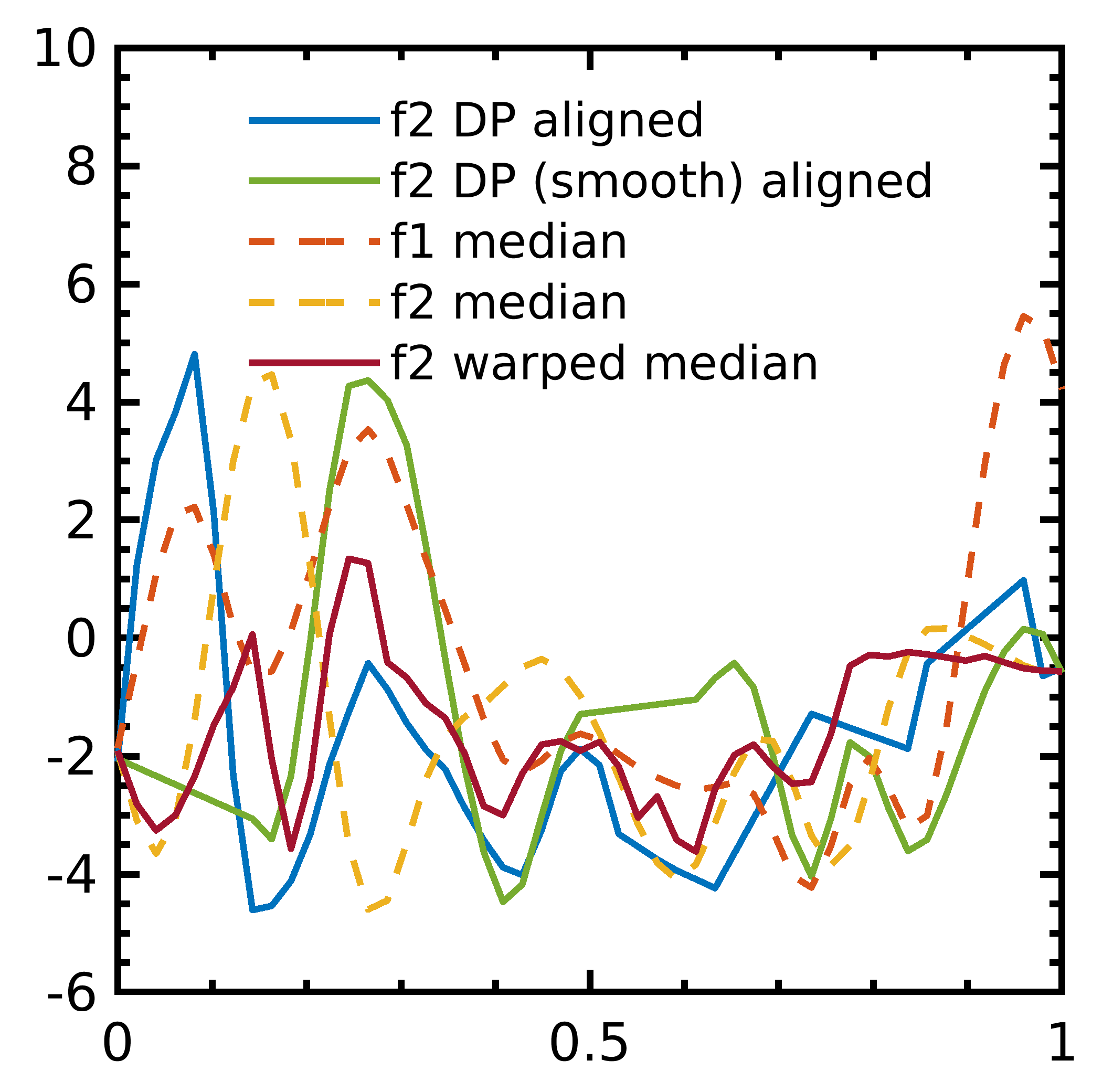}
\caption{Comparison of the posterior distribution $\gamma|y_1,y_2$ to the DP solution $\gamma_{DP}$ applied to the iPhone data, plotting the two identified modes and respective 95\% credible regions of the posterior $\gamma|y_1,y_2$ (left). Comparison of the DP warping of $y_2$ to the median of the warping $f_2 \circ \gamma|y_1,y_2$ (right).}
\label{fig:iphone_gampost}
\end{figure}

The posteriors $f_1|y_1$ and $f_2|y_2$ can be seen in the left panel of Figure \ref{fig:iphone_f1f2post} along with $y_1$ and $y_2$ and the medians are shown in the right panel of Figure \ref{fig:iphone_gampost}. 
This image also compares the DP aligned solution $(y_2 \circ \gamma_{DP})$ to the posterior alignment of $f_2|y_2$, $(f_2 \circ \gamma) |y_1,y_2,$ from our Bayesian approach. It is clear that our Bayesian approach can capture multiple possible alignments, including alignments similar to the DP solution. For example, in addition to the alignment captured by the DP algorithm, the Bayesian approach also accounted for an alignment of $f_2$ to the leftmost peak of $f_1$.
It is interesting to note here that the method didn't match $f_2$ to the rightmost peak of $f_1$. This would be caused by an extremely distorted warping function and the Fisher Rao metric natively guards against that by penalizing too high of a gradient. 

Figure \ref{fig:iphone_gampost} on the left compares the DP solution to the posterior distribution of $\gamma|y_1,y_2$ and shows the posterior median and $95\%$ credible region of $\gamma|y_1, y_2$. 
The credible region quantifies the uncertainty around the estimated warping function $\gamma$ and warped function $(f_2 \circ \gamma)$.
The DP solution, obtained by trying to align the noisy data without prior data smoothing, is not built to account for the multiple possible alignments. 

\section{Discussion}\label{sec:disc}

We have proposed a new flexible Bayesian approach to functional alignment that handles measurement error and accounts for multiple optimal alignments in the posterior of the warping function $\gamma.$ 
We have demonstrated its advantages over the state-of-the-art Dynamic Programming method that has been used in recent alignment literature \cite{FSDA,srivastava2011,tucker2013} using both simulated and real datasets. 
Unlike DP, a Bayesian approach can characterize the uncertainty in the warping function solution $\gamma$, with Bayesian credible intervals.
The hierarchical structure of our Bayesian method allows us to estimate measurement error observed in $y_1$ and $y_2$ and extract the mean functions $f_1$ and $f_2$ to be aligned, respectively. 

Accounting for measurement error directly in the method itself avoids any need for smoothing of the data typically done prior to applying DP in practice.
Additionally, by running parallel MCMC chains, the posterior of the warping function $\gamma$ is able to capture all possible alignments and visualize the posterior modes and respective credible regions for one or more distinct alignments.
Although existing Bayesian methods can account for uncertainty in the warping function, they are not able to find more than one possible alignment.
This is because they focus on converging efficiently to one solution, and do not account for measurement uncertainty.
Without accounting for all possible alignments, we are missing information; and not accounting for all uncertainty in $\gamma$.

For future work, we look to extend the method to the multiple alignment problem. 
This will not be trivial in the case of multi-modal posteriors and multi-function alignment, as the solution will not be as simple as running multiple chains. 
The extension in our case would be the following: given $I$ functions you would construct $I$ pairwise alignments in Level 1 and Level 2 and then model the mean or template function using a Gaussian process prior and update. 
This processes would be somewhat similar to what was proposed in \cite{lu2017}.
However, there lies the additional difficulty in sampling a model which can give multiple solutions. For some cases, each pairwise alignment to the template would produce multiple possible gamma functions. 
One would either need to determine which mode (e.g., \textit{optimal}) warping function to use or the uncertainty would need to be propagated to the sampling of the mean function. 
As noted in Cheng et al., \cite{cheng2016}, this template is only identifiable up to an equivalence class of warpings. 
The issue is more pronounced in the case of multiple alignments.
The use of wormhole MCMC (\cite{lan2014}) or something similar would need to be explored and possibly used in this situation. 
Additionally, we can extend this from curves to trajectories that lie on Riemannian manifolds $\mathcal{M}$.
In this case, one has to account for the non-zero curvature of the space and in particular, the calculation of the gradient in the Fisher Rao metric.

\section*{Acknowledgment}
This paper describes objective technical results and analysis. Any subjective views or opinions that might be expressed in the paper do not necessarily represent the views of the U.S. Department of Energy or the United States Government. This work was supported by the Laboratory Directed Research and Development program at Sandia National Laboratories; a multi-mission laboratory managed and operated by National Technology and Engineering Solutions of Sandia, LLC, a wholly owned subsidiary of Honeywell International, Inc., for the U.S. Department of Energy's National Nuclear Security Administration under contract DE-NA0003525.

\bibliography{biblio} 
\bibliographystyle{elsarticle-harv}

\appendix
\section{Derivative of $\Phi(g)$}\label{sec:deriv}

In this section we provide the derivations to compute the derivative of the negative log-likelihood, $\nabla\Phi(\psi)=\frac{d\Phi}{d\psi}$. The negative log-likelihood is defined

$$\Phi:=-\log L(\psi,\sigma_1^2) = \frac{N}{2}\log (\sigma_1^2)+ \frac{1}{2\sigma_1^2}\int_0^1\left(q_1(t)- (q_2(\int_0^t\psi^2(s)\,ds)\psi(t))\right)^2\,dt.$$

To compute the directional derivative, let $A(t) = q_1(t)- (q_2(\int_0^t\psi^2(s)\,ds)\psi(t))$ and we will rewrite $\nabla \Phi(\psi)$ for simplicity as

$$\nabla \Phi(\psi)  = \frac{1}{\sigma_1^2}\int_0^1 A(t)\nabla_\psi A(t)\,dt.$$

To find the directional derivative $\nabla_\psi A(t_i)$, we first,
consider the sequence of maps $\psi \overset{\int_0^t \psi^2 ds}{\mapsto}
\gamma \overset{\phi}{\mapsto} r $, where $r := \phi(\gamma) =
(q\circ\gamma)\sqrt{\dot{\gamma}}$. For the constant function $\1\in\Psi$ and
a tangent vector $u\in T_\1(\s_{\infty})$ the differential of the first
mapping at $\1$ is $2 \bar{u}(t) = 2 \int_0^t u(s) ds$. For a tangent vector
$w\in T_{\gamma_{id}}(\Gamma)$, the differential of the second mapping at
$\gamma_{id} = t$ is $\frac{\partial \tilde{q}}{\partial t} w +
\frac{1}{2}\tilde{q}\dot{w}$, where $\tilde{q} = q(\int_0^t \psi(s)^2
ds)\psi$. If we concatenate these two linear maps we obtain the directional
partial derivative of $A(\psi)$ in a direction $u\in
T_\1(\s_{\infty})$ as
\[\nabla_\psi A(u) =  -2\frac{\partial \tilde{q}_2}{\partial t} \bar{u}(t)-\tilde{q}_2 u(t).\]

We now can write the derivative of $\Phi$ in the direction of $u$ as
\[\nabla_\psi \Phi(u)  = \frac{1}{\sigma_1^2}\int_0^1 A(t)\nabla_\psi A(u)\,dt.\]

Since $T_\1(\s_{\infty})$ is an infinite-dimensional space, we can approximate
the directional partial derivative by considering a finite-dimensional
subspace of $T_\1(\s_{\infty})$. Let us form a subspace of $T_\1(\s_{\infty})$
using $\{(\frac{1}{\sqrt{\pi}}\sin(2\pi n t),\frac{1}{\sqrt{\pi}}\cos(2\pi n
t))\mid n=1,2,\dots,p/2\}$. We then can approximate the derivative using
\[ \nabla_\psi \Phi = \sum_{k=1}^p \nabla_\psi \Phi(c_k) c_k,\]
where $c_k$'s are the $k=1,\ldots,p$ basis elements of the subspace.

\section{MCMC diagnostics}\label{sec:mcmcdiag}

In this section, we show select MCMC diagnostic plots for the three examples demonstrated in this paper. We show the accepted samples for $\sigma_1^2$ and $\sigma_2^2$ for all chains run in parallel which show the good mixing and convergence of $f_1|y_1$ and $f_2|y_2.$ We also show the accepted samples for $\sigma^2$ for each chain separately to show convergence of $(q_1- \mathcal{G} (g(t)))|y_1,y_2$ for each chain. Lastly, we show the accepted samples for the $n_v$ basis coefficients $B$ for all chains to show their ability to jump between modes and mix well within a mode. Note that due to the high efficiency of the $\infty$-HMC algorithm and necessary burn-in to ensure all parameters converge, we expect the relatively low acceptance rates for $(g,\theta)|y_1,y_2$ observed in the right hand figure of Figures \ref{fig:simex_chains}, \ref{fig:sonar_chains}, and \ref{fig:iphone_chains} below.

\begin{figure}[H]
\centering
\includegraphics[width=0.32\textwidth]{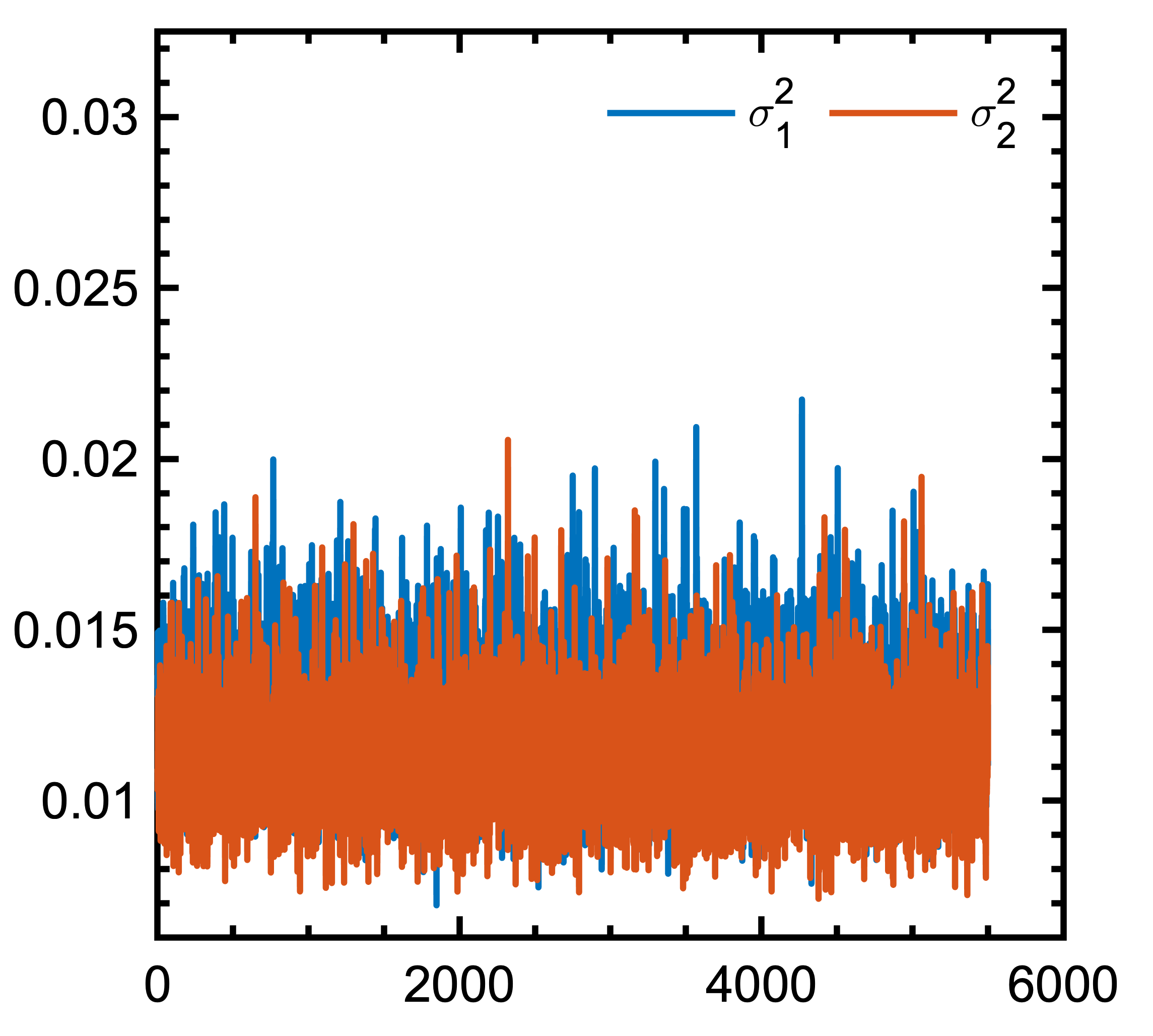}
\includegraphics[width=0.33\textwidth]{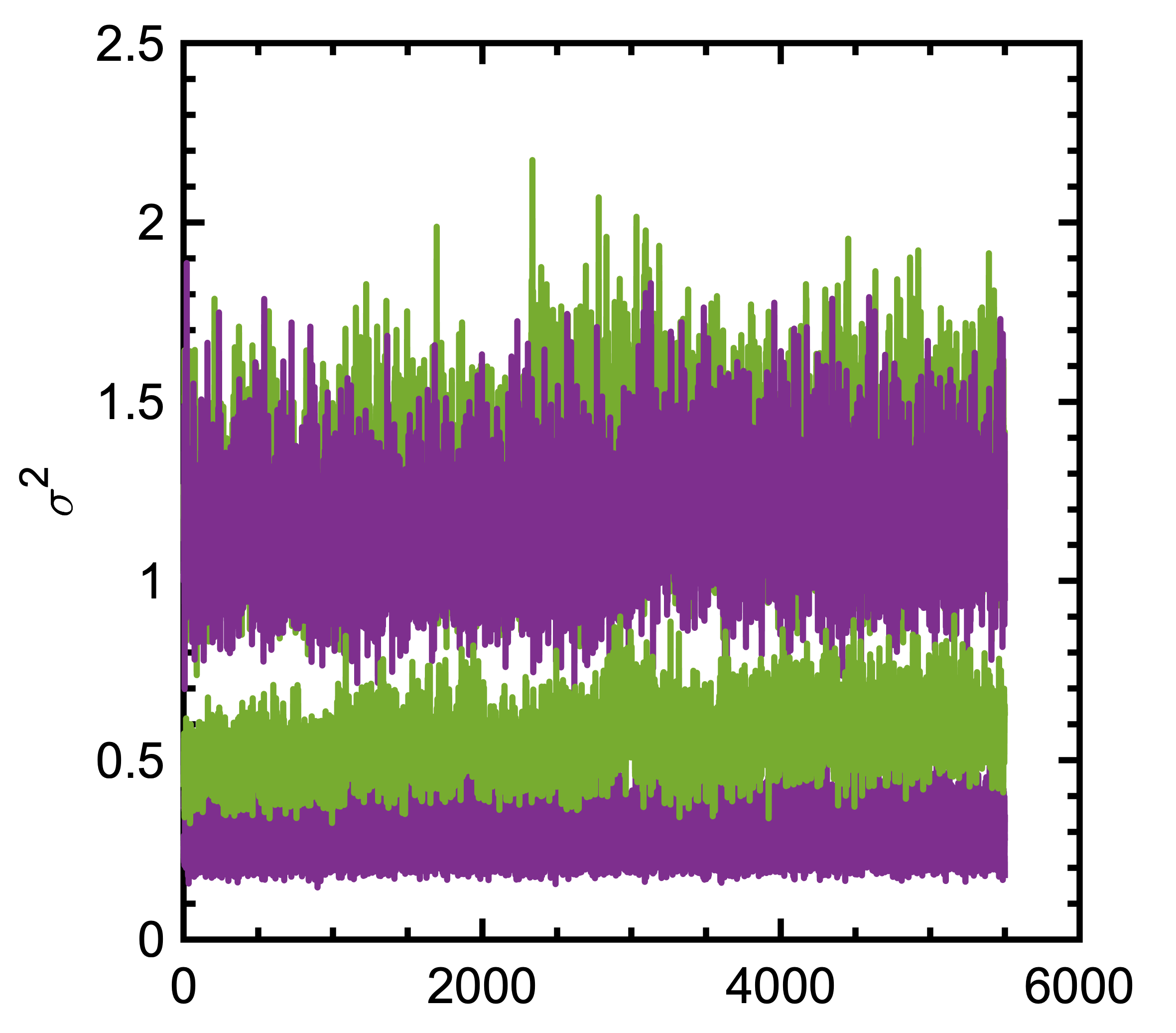}
\includegraphics[width=0.32\textwidth]{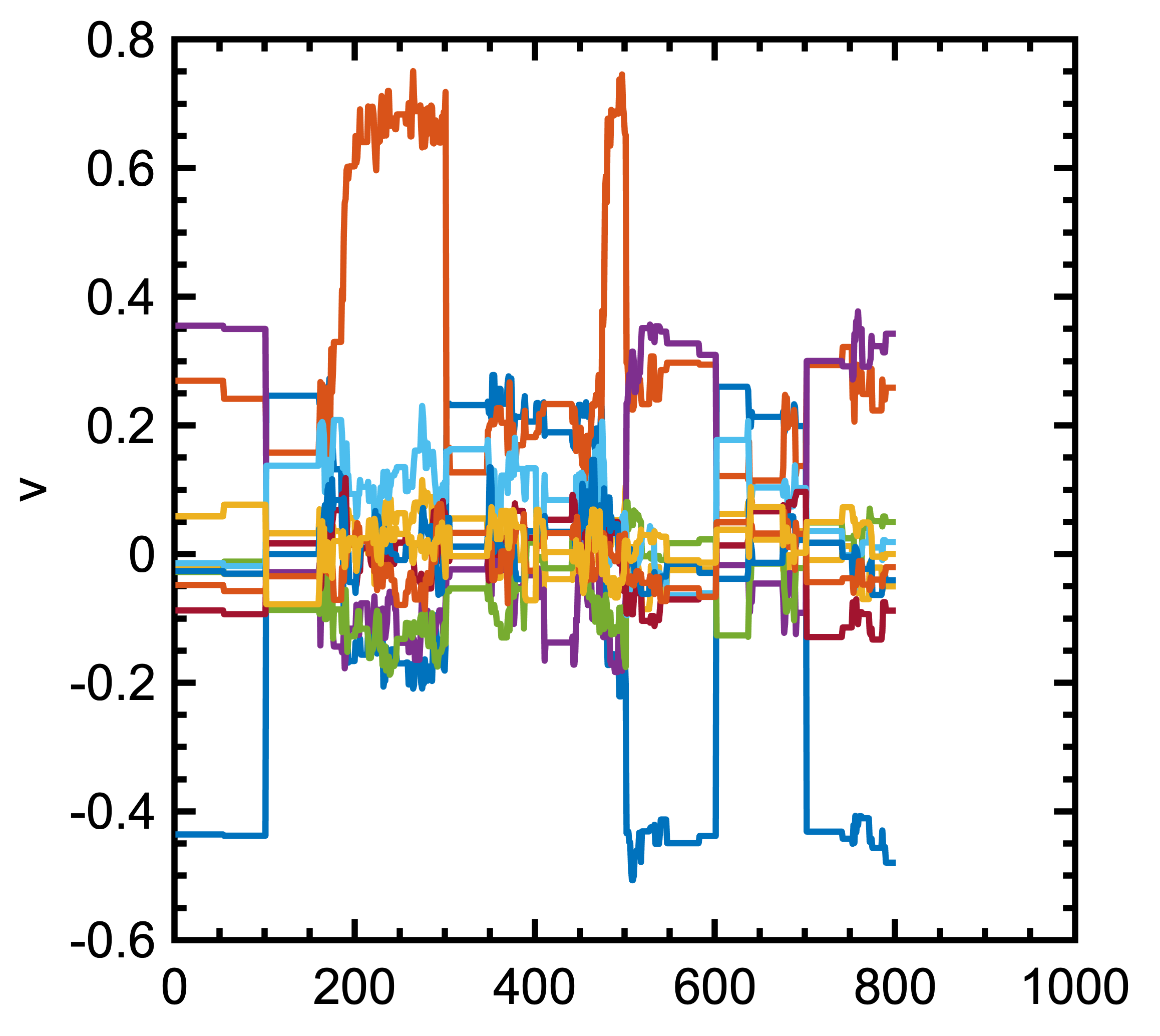}
\caption{From left to right: MCMC chains from the simulated data results shown in Section \ref{sec:simex}  for $\sigma_1^2$ (blue), $\sigma_2^2$ (orange), $\sigma^2$ for all 8 chains separately colored by the posterior cluster they are associated with, and $\bf{v}$ for all chains).}
\label{fig:simex_chains}
\end{figure}

\begin{figure}[H]
\centering
\includegraphics[width=0.32\textwidth]{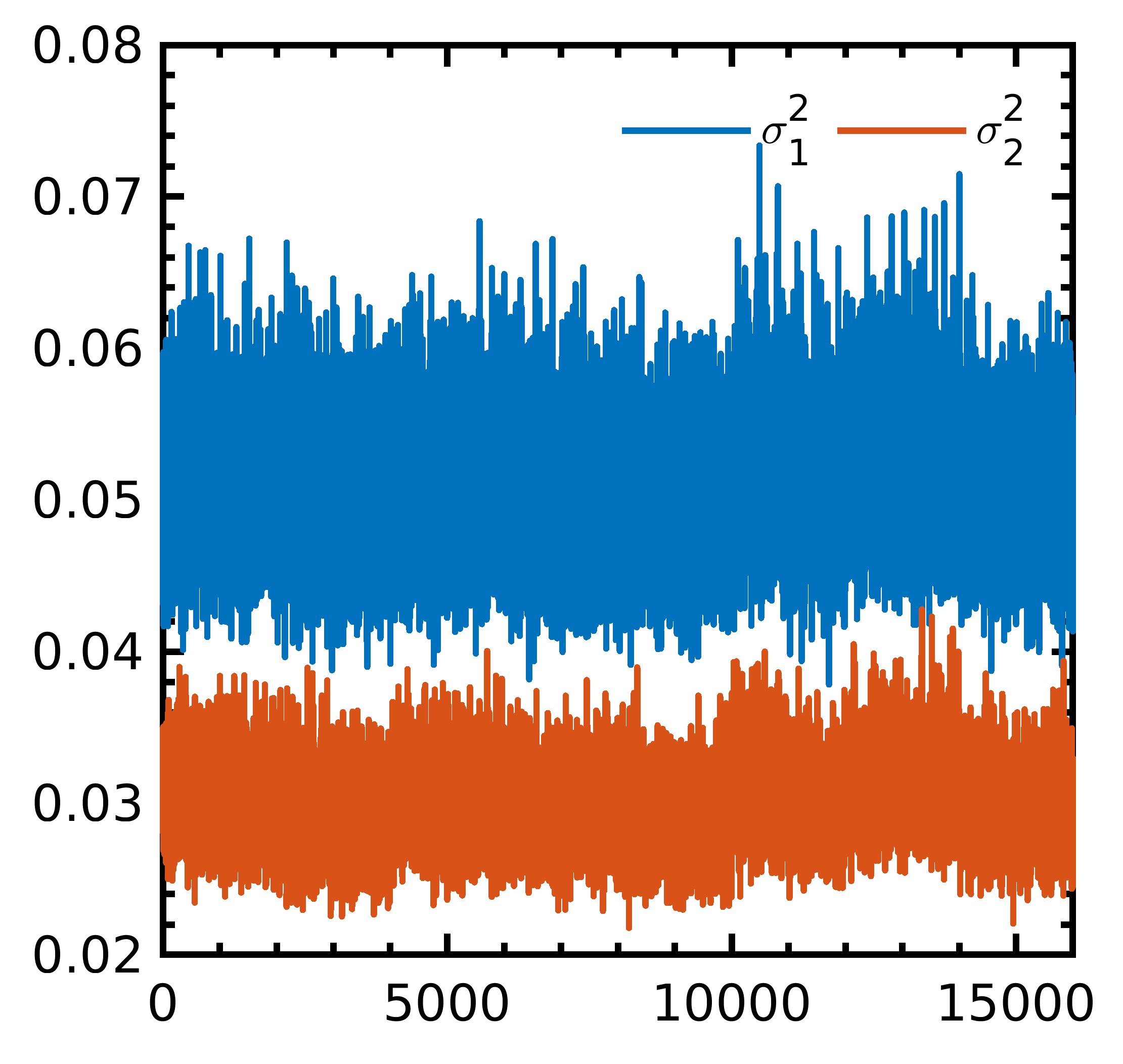}
\includegraphics[width=0.335 \textwidth]{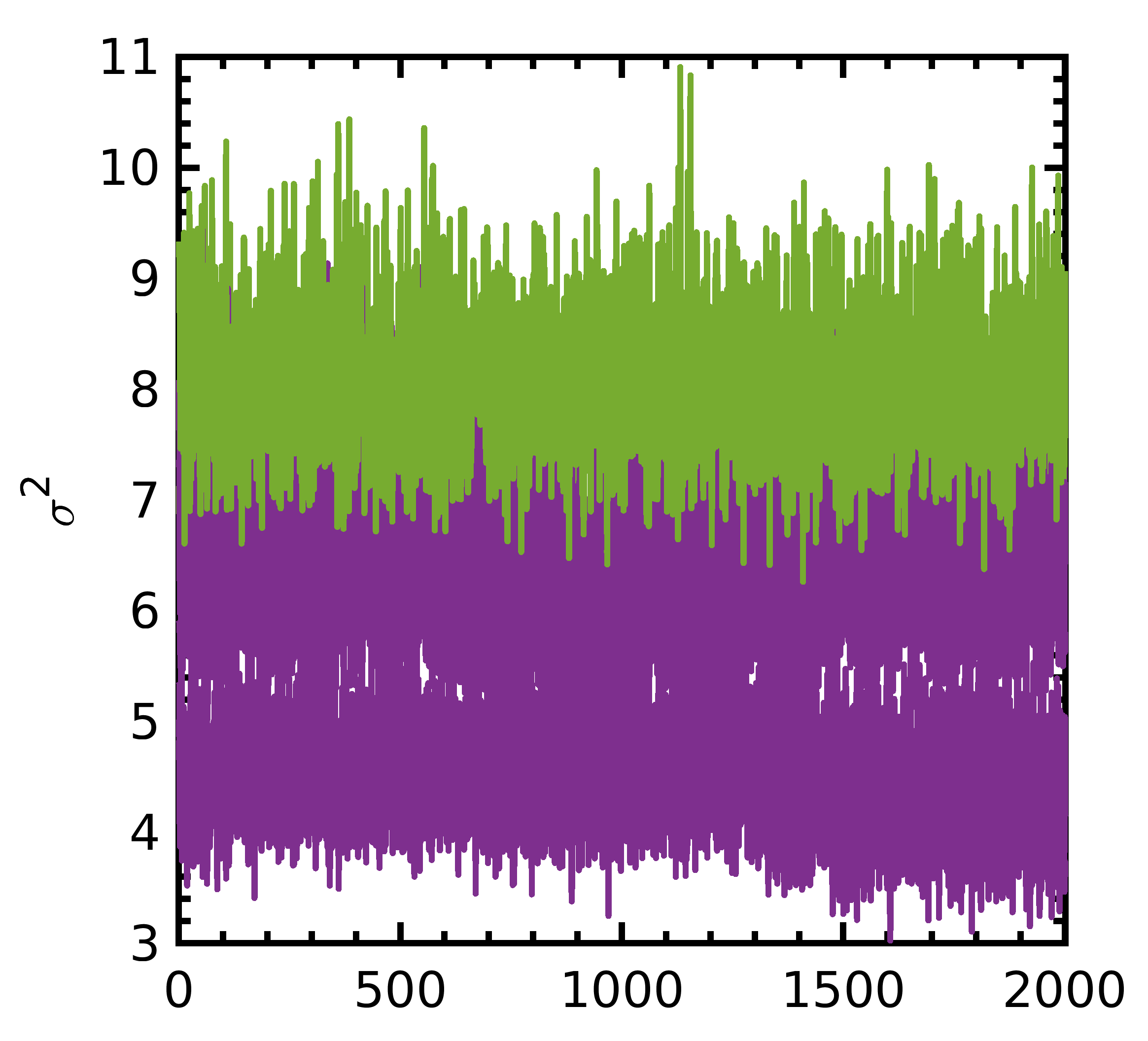}
\includegraphics[width=0.325\textwidth]{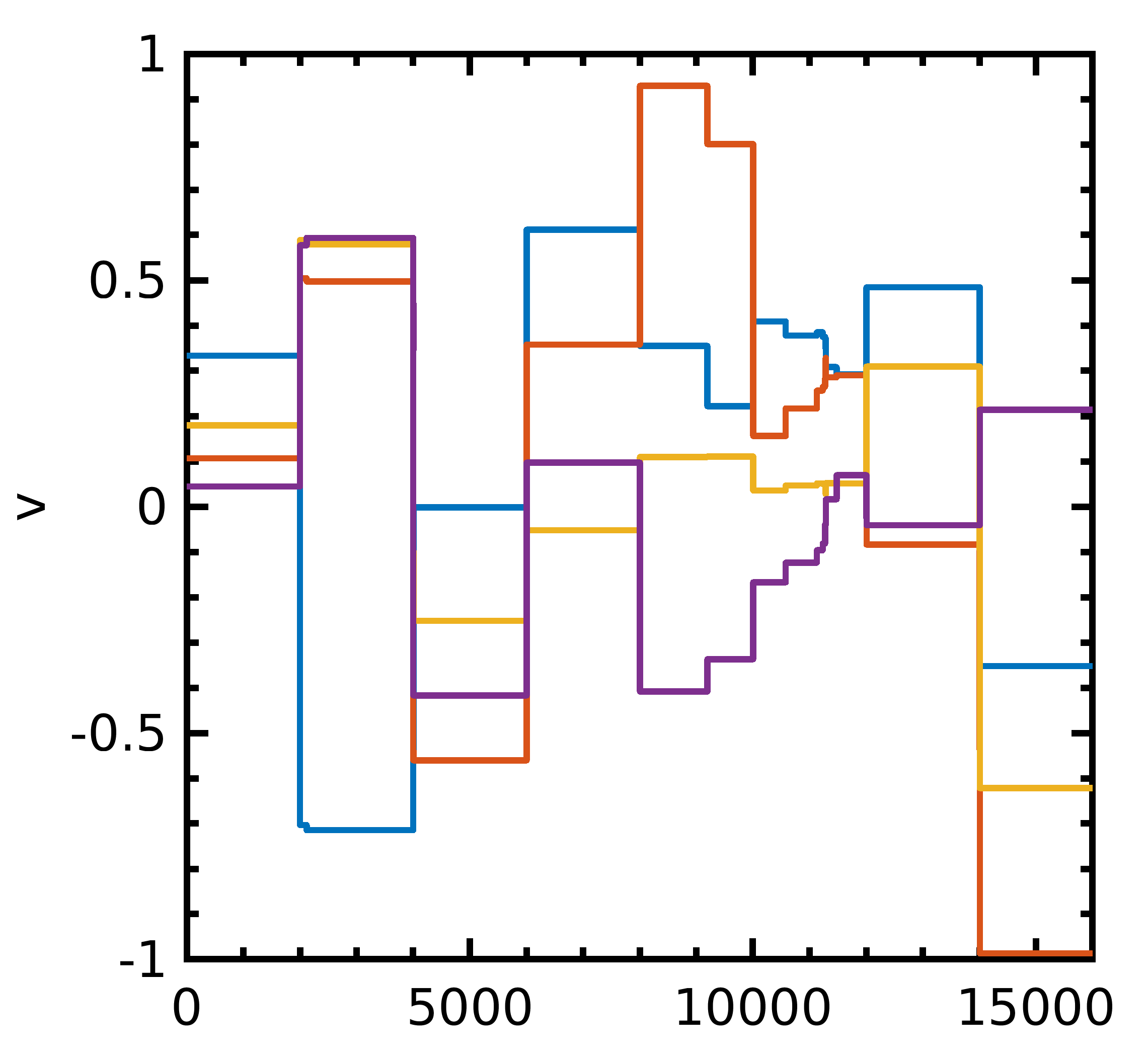}
\caption{From left to right: MCMC chains from the SONAR data results shown in Section \ref{sec:iphone}  for $\sigma_1^2$ (blue), $\sigma_2^2$ (orange), $\sigma^2$ for all 8 chains separately (middle) colored by the posterior cluster they are associated with, and $\bf{v}$ for all chains.}
\label{fig:sonar_chains}
\end{figure}

\begin{figure}[H]
\centering
\includegraphics[width=0.32\textwidth]{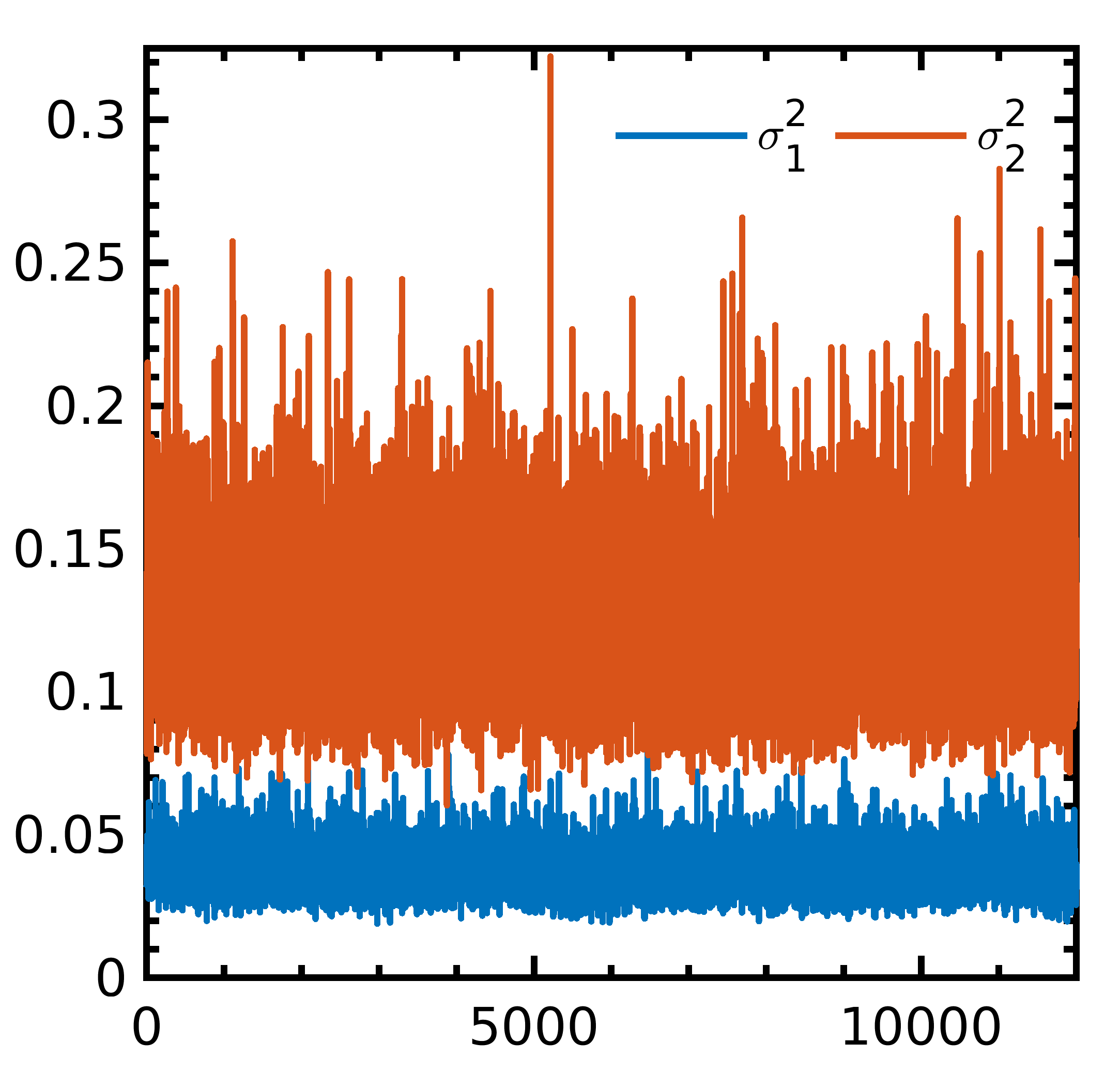}
\includegraphics[width=0.335\textwidth]{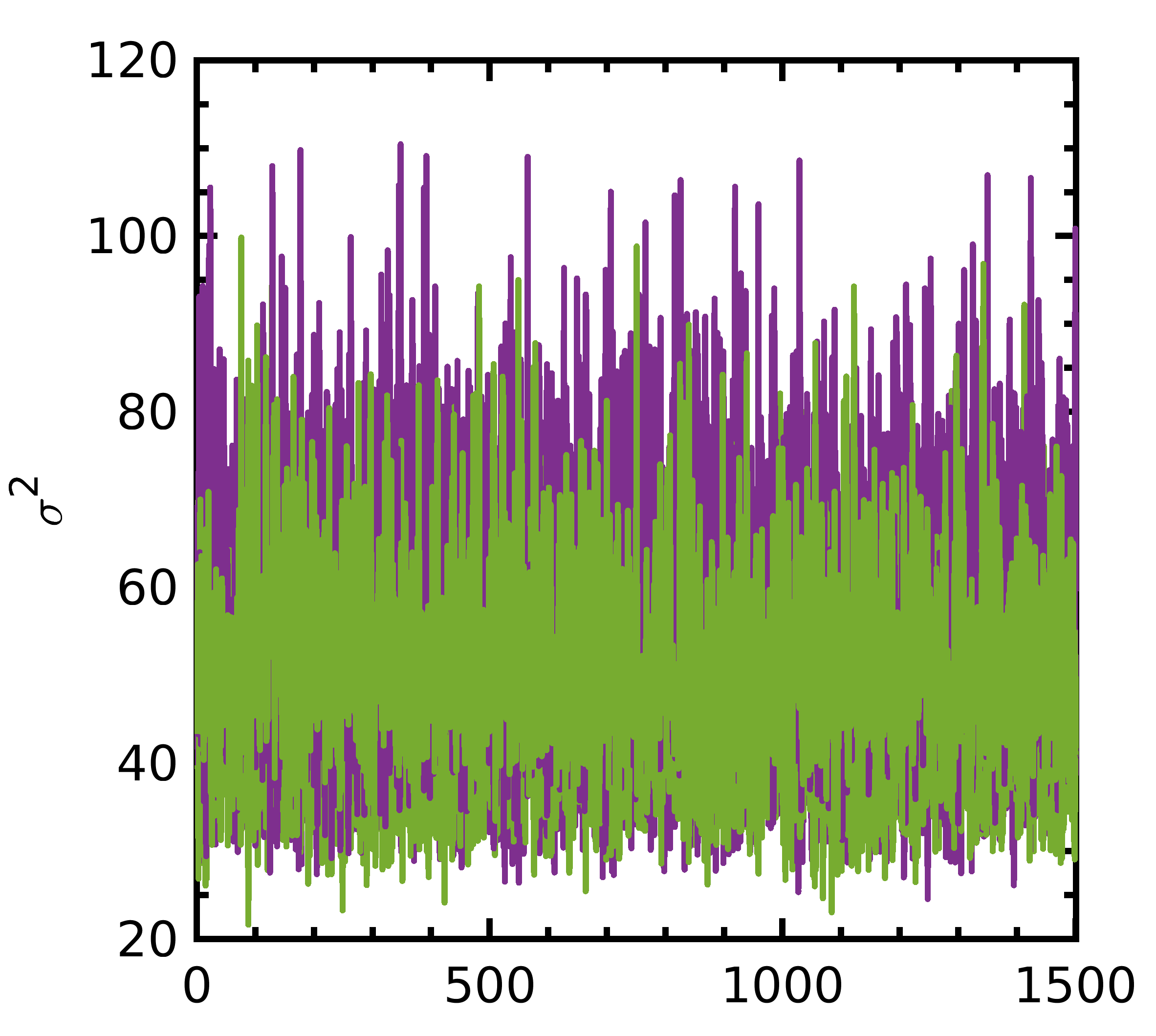}
\includegraphics[width=0.325\textwidth]{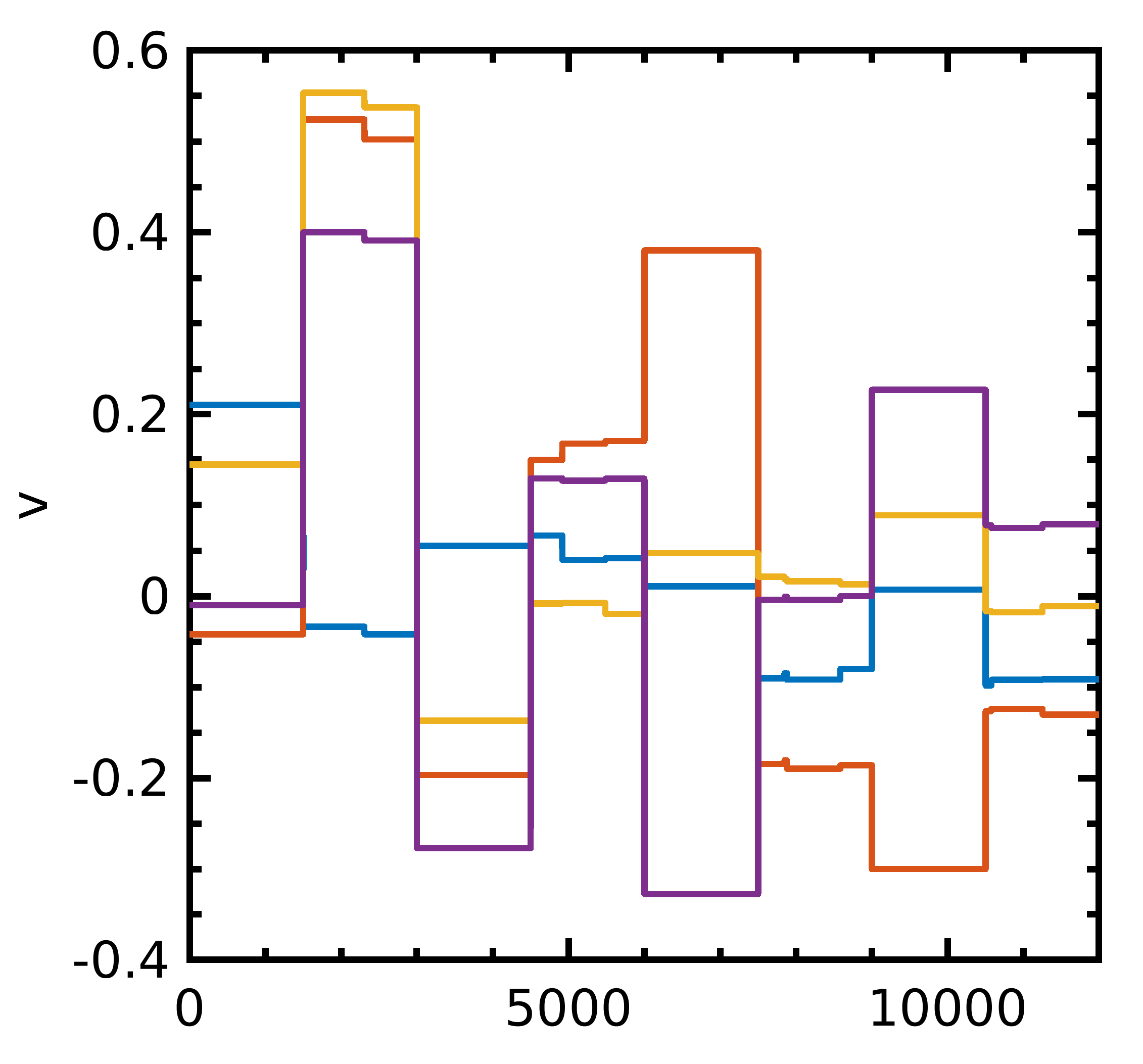}
\caption{From left to right: MCMC chains from the iPhone data results shown in Section \ref{sec:iphone}  for $\sigma_1^2$ (blue), $\sigma_2^2$ (orange), $\sigma^2$ for all 8 chains separately colored by the posterior cluster they are associated with, and $\bf{v}$ for all chains.}
\label{fig:iphone_chains}
\end{figure}

\begin{figure}[H]
  \centering
  \includegraphics[width=0.47\textwidth]{bimodal_gammaall.png} 
  \includegraphics[width=0.49\textwidth]{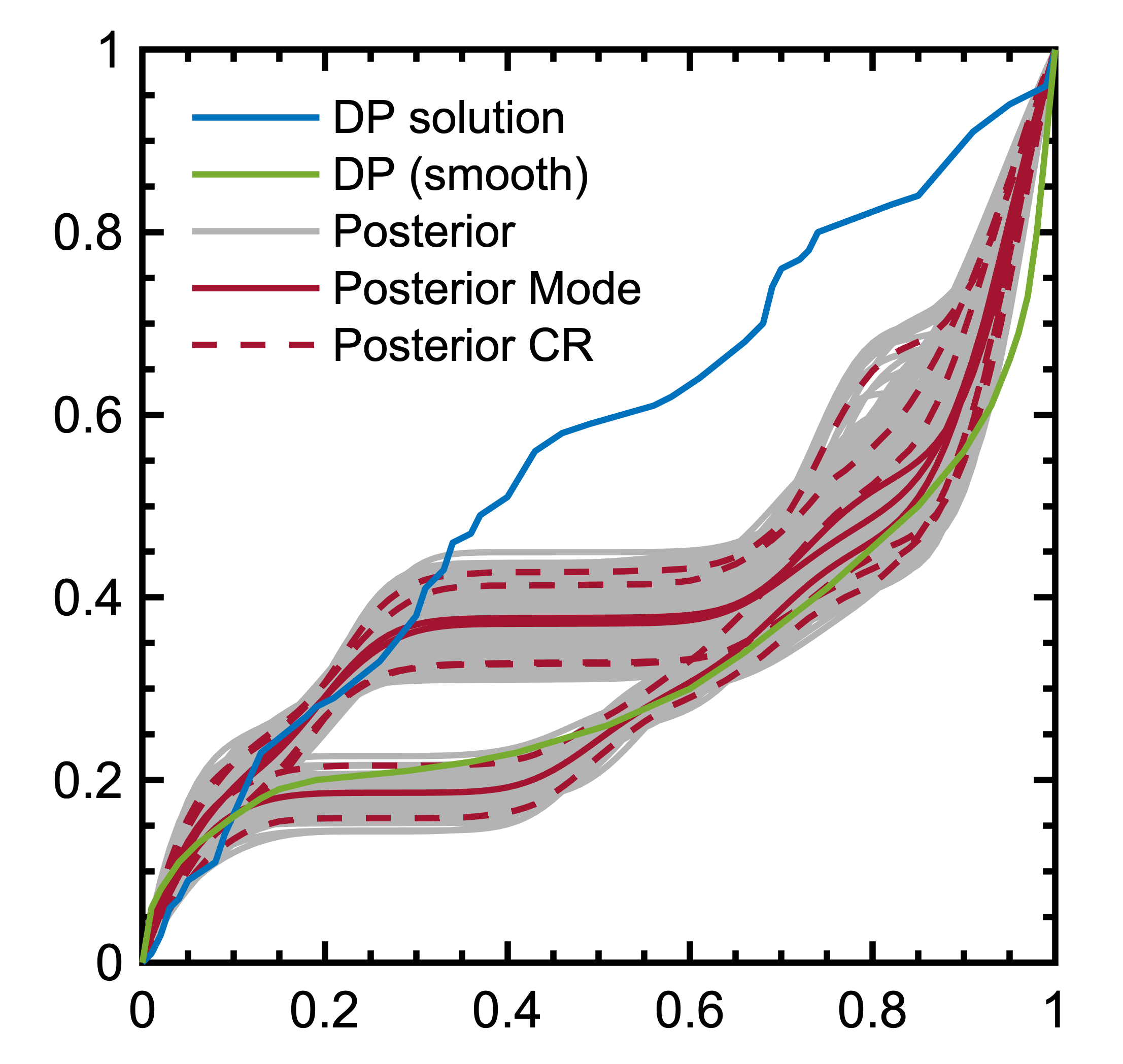}
  \caption{Comparison of the posterior distribution $\gamma|y_1,y_2$ to the DP solution $\gamma_{DP}$ applied to the simulated data using $\infty$-HMC (left) and Z-mixutre pCN (right).}
  \label{fig:comparison_samplers}
\end{figure}

\begin{figure}[H]
  \centering
  \includegraphics[width=0.67\textwidth]{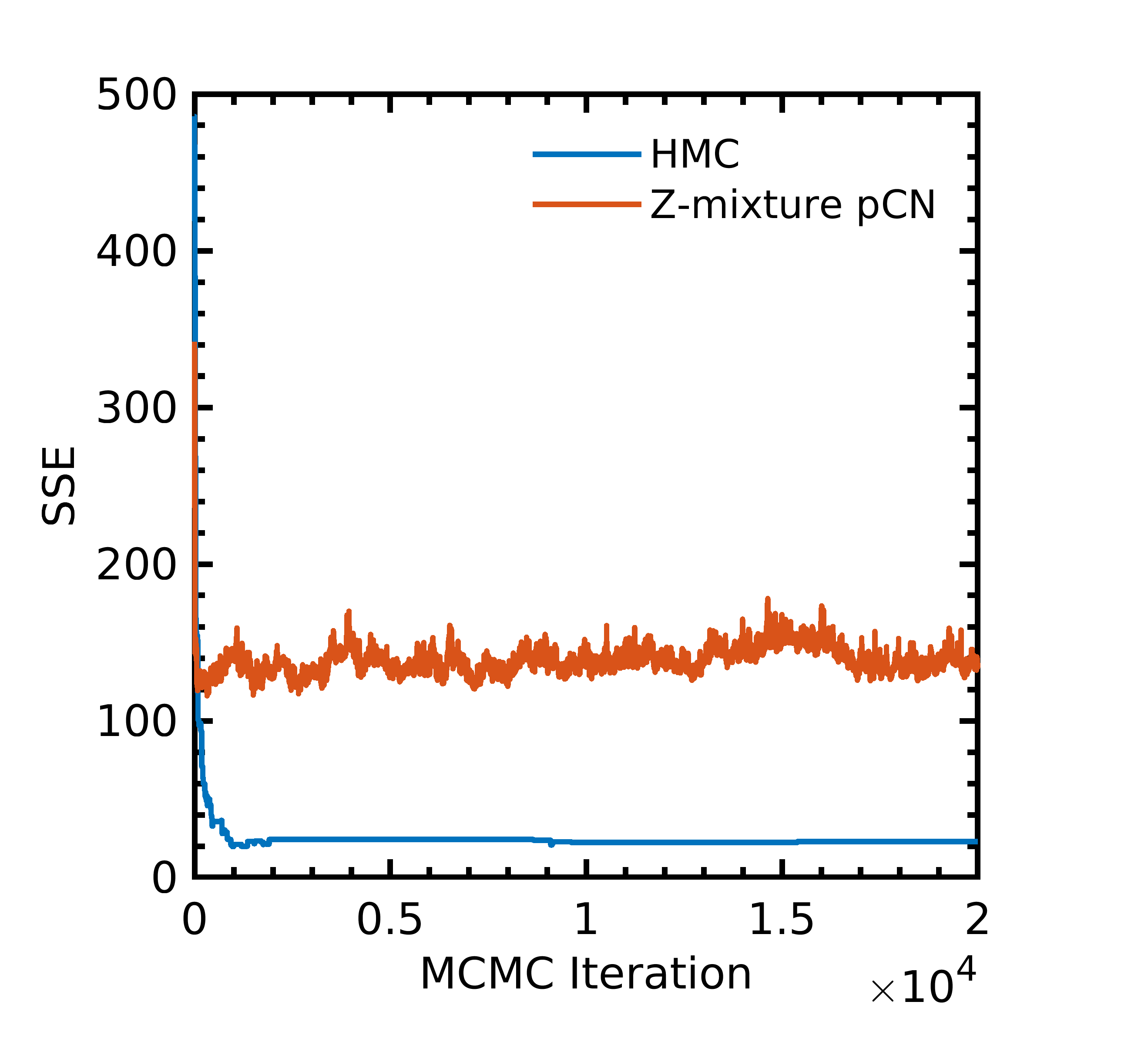} 
  \caption{Sum-of-squared errors (SSE) for the $\infty$-HMC and Z-mixture pCN samplers.}
  \label{fig:comparison_SSE}
\end{figure}

Figure~\ref{fig:comparison_samplers} provides a comparison of the posteriors when using $\infty$-HMC (left panel) and the Z-mixture pCN algorithm (right panel) on the simulated data.
Both samplers were optimized for the data and we used 8 chains for 5000 iterations. 
Each of the samplers are able to find two of the modes, but the Z-mixture pCN fails to find one of the modes.
Since the Z-mixture pCN is a random walk type algorithm and we are sampling from an infinite dimensional space, it gets stuck where it is started and can fail to move outside the local area it is in. 
Additionally, we compared the SSE values for the two samples for two chains that give a similar warping function. 
Figure~\ref{fig:comparison_SSE} provides the SSE values and the $\infty$-HMC achieves a lower SSE for similar solutions and this is consistent across multiple chains. 

\end{document}